\documentclass[aps,prd,a4paper,preprintnumbers,tightenlines,superscriptaddress,
twocolumn,floatfix,showpacs,final,nofootinbib]{revtex4-1}
\usepackage{graphicx}
\usepackage{longtable}
\usepackage{amsmath}
\usepackage{amssymb}
\usepackage{subfigure}
\usepackage{hyperref}
\usepackage{dcolumn}
\usepackage{bm}
\usepackage{color}

\newcommand{\dd}{\, {\rm d}}
\newcommand{\gsim}{\;\mbox{\raisebox{-0.5ex}{$\stackrel{>}{\scriptstyle{\sim}}$}
}\;}
\newcommand{\lsim}{\;\mbox{\raisebox{-0.5ex}{$\stackrel{<}{\scriptstyle{\sim}}$}
}\;}

\newcommand{\pn}{\Phi_{\rm N}}

\newcommand{\kar}{\kappa_{\rm R}}

\newcommand*\colvec[3][]{\begin{pmatrix}\ifx\relax#1\relax\else#1\\\fi#2\\#3\end{pmatrix}}

\DeclareMathOperator{\sinc}{sinc}

\newcommand{\ccc}{{_{\rm c}}}

\newcommand{\eee}{_{\rm e}}

\newcommand{\GN}{G_{\rm N}}

\def\eea{\end{eqnarray}}
\def\bea{\begin{eqnarray}}
\allowdisplaybreaks
\usepackage{enumitem}
\setenumerate[1]{label=(\arabic*)}

\begin{document}
\title{Testing Gravity Using Dwarf Stars}
\author{Jeremy Sakstein}
\email[Email:]{jeremy.sakstein@port.ac.uk}
\affiliation{Institute of Cosmology and Gravitation,
 University of Portsmouth, Portsmouth PO1 3FX, UK}

\begin{abstract}
Generic scalar-tensor theories of gravity predict deviations from Newtonian physics inside astrophysical bodies. In this paper, we point out 
that low mass stellar objects, red and brown dwarf stars, are excellent probes of these theories. We calculate two important and 
potentially observable quantities: the radius of brown dwarfs and the minimum mass for hydrogen burning in 
red dwarfs. The brown dwarf radius can differ significantly from the GR prediction and upcoming surveys that probe the mass-radius relation for stars 
with masses $<\mathcal{O}(0.1M_\odot)$ have the potential to place new constraints. The minimum mass for hydrogen burning can be larger than several 
presently observed Red Dwarf stars. This places a new and extremely 
stringent constraint on the parameters that appear in the effective field theory of dark energy and rules out several well-studied dark energy models.
\end{abstract}
\maketitle

\section{Introduction}

Dark energy and the cosmological constant problem have been driving the study of alternative theories of gravity for more than a decade (see 
\cite{Clifton:2011jh,Joyce:2014kja} for recent reviews). Theories that are able to drive the acceleration of the cosmic expansion typically invoke 
\textit{screening mechanisms} in order to remain compatible with local tests of gravity. These use non-linear features to decouple solar system and 
cosmological scales such that the theory behaves like general relativity (GR) in the solar system whilst exhibiting novel cosmologies. This has 
prompted a large theoretical effort towards finding novel small scale probes of these mechanisms and astrophysical probes, in particular stellar 
structure tests, have emerged as promising candidates 
\cite{Chang:2010xh,Davis:2011qf,Jain:2012tn,Brax:2013uh,Sakstein:2013pda,Vikram:2014uza,Sakstein:2014nfa,Sakstein:2015oqa,Koyama:2015oma,
Sakstein:2015aqx}. 

The Vainshtein mechanism \cite{Vainshtein:1972sx} is very efficient at hiding deviations from GR locally and is exhibited in a wide variety of 
alternative gravity theories \cite{Babichev:2009us,Kimura:2011dc,Koyama:2013paa}. Recently, \cite{Kobayashi:2014ida} have shown that the mechanism is 
partially broken inside astrophysical bodies in a wide class of interesting theories (the beyond Horndeski class 
\cite{Gleyzes:2014dya,Gleyzes:2014qga}). Using this result, \cite{Koyama:2015oma} have derived the modifications to the stellar structure 
equations\footnote{See also \cite{Saito:2015fza}.}, which manifest as a modification of the hydrostatic equilibrium equation. This was applied to 
polytropic models to derive the new Lane-Emden equation, which was used to predict the properties of main-sequence 
stars. The breaking of the mechanism can result in weaker gravity inside extended objects and this results in stars that are dimmer and cooler at 
fixed mass\footnote{There is a small region of parameter space where the converse is true \cite{Saito:2015fza}. We will discuss this later.}. 
Polytropic models are an incomplete model of main-sequence stars and lack a description of nuclear burning and metallicity effects. As such, 
\cite{Koyama:2015oma} identified several degeneracies that may make the predicted novel effects difficult to observe in practice.

In this paper, we point out that low mass stellar objects, red and brown dwarf stars, are perfect probes of these theories precisely because they are 
well-described by polytropic models. They are chemically homogeneous except for the photosphere and their observational properties are only weak 
functions of their opacity. Furthermore, the lack of chemical evolution (except during their very early phases) allows us to treat them as static.

Low mass brown dwarfs have a radius that is almost independent of their mass over a large range of masses. Polytropic models based on GR predict that 
this is $0.1R_\odot$, which is close to the observed value. In this work, we find that this can be arbitrarily large when gravity is weaker, 
and as small as $0.078R_\odot$ when it is stronger. Measuring the mass-radius relation of objects with mass $M\lsim\mathcal{O}(0.1M_\odot)$ can 
therefore place a new constraint on alternative gravity theories. 

Higher mass objects obey a different mass-radius relation whereby they are more compact at 
higher masses. Objects that are massive enough have sufficiently high core temperatures and densities to burn hydrogen. Thus, only stars that are 
sufficiently massive can burn on the 
main-sequence; lower mass objects remain inert and eventually cool. GR predicts that the minimum mass for hydrogen burning (MMHB) is around 
$0.08M_\odot$, which is again very close to what is observed. Here, we find that alternative gravity models predict values that can be arbitrarily 
large when gravity is weaker\footnote{We will not consider the case where gravity is stronger because this is always consistent with observations of 
the lowest mass hydrogen burning objects.}. The observation of several hydrogen burning stars with masses in the 
range $0.08$--$0.1M_\odot$ then allows us to constrain the new parameter $\Upsilon\lsim 1.4$. In terms of the effective field theory of dark 
energy, this places an independent constraint on the parameters
\begin{equation}
 \frac{\alpha_H^2}{\alpha_H-\alpha_T-\alpha_B(1+\alpha_T)}\lsim 0.4.
\end{equation}
This constraint was presented in a recent letter \cite{Sakstein:2015zoa}. Here, we expand the calculation and discuss it in detail.

This paper is organised as follows: In section \ref{sec:vain} we provide an introduction to the Vainshtein mechanism for readers unfamiliar with 
screening mechanisms and alternative theories of gravity. We also present the modification of the hydrostatic equilibrium equation and apply it to 
polytropic models to derive the new Lane-Emden equation. The connection with the effective field theory of dark energy is also presented. In section 
\ref{sec:BD} we provide a brief review of the salient features of dwarf 
stars for the benefit of readers well-versed in alternative gravity theories who may be unfamiliar with stellar astrophysics. In section 
\ref{sec:MGBD} 
we apply the modified Lane-Emden equation to dwarf star models in order to calculate the alternative gravity theory predictions for their 
observable properties. We conclude in section \ref{sec:concs}.

\section{alternative Gravity Theories and the Vainshtein Mechanism}\label{sec:vain}

\subsection{The Vainshtein Mechanism}

The Newtonian limit of GR results in a scalar theory of gravity where the gravitational field is described by the Newtonian potential $\pn$, which 
results in a force (per unit mass) $F_{\rm N}=-\nabla\pn$. The gravitational field is sourced by the local density $\rho$ through the Poisson equation
\begin{equation}
\nabla^2\pn=4\pi G\rho.
\end{equation}
Scalar-Tensor alternatives to GR include an additional field $\phi$, which results in a \textit{fifth-force}
\begin{equation}
F_5=-\beta\nabla\phi
\end{equation}
parametrised by a dimensionless coupling $\beta\sim\mathcal{O}(1)$. The simplest scalar-tensor theories generically predict that the local field is 
sourced by the density in the exact same manner as the Newtonian potential \cite{EspositoFarese:2004cc}:
\begin{equation}\label{eq:ststand}
\nabla^2\phi=8\pi\beta G\rho
\end{equation}
so that one has $\phi=2\beta\pn$. The total force is then $F=(1+2\beta^2)F_{\rm N}$ and one must typically tune $\beta\lsim 10^{-3}$ to satisfy 
solar system bounds \cite{EspositoFarese:2004cc}. Importantly, 
one then has $\beta\ll1$ on all scales, which makes it difficult for $\phi$ to have any non-trivial influence on the cosmology. 

The Vainshtein mechanism circumvents this by introducing new derivative interactions that manifest as new differential operators on the left hand 
side of (\ref{eq:ststand}). This results in solutions for $ \phi$ that differ significantly from $\pn$. One of the simplest examples of this is the 
cubic galileon model \cite{Nicolis:2008in}, which, imposing spherical symmetry, results in the following equation:
\begin{equation}
\frac{1}{r^2}\frac{\dd }{\dd r} \left(r^2\frac{\dd\phi}{\dd r}\right)+\frac{1}{r^2}\frac{\dd }{\dd r}\left[r\left(\frac{\dd\phi}{\dd 
r}\right)^2\right]=8\pi \beta G\rho.
\end{equation}
The first term can be recognised as $\nabla^2\phi$; the second is the new term arising from the new interaction. Integrating once, we obtain an 
algebraic relation for the ratio of the fifth- to Newtonian force $x\equiv F_5/F_{\rm N}$:
\begin{equation}
x+\left(\frac{r_{\rm V}}{r}\right)^3\frac{x^2}{2\beta^2}=2\beta^2,\quad\textrm{where}\quad r_{\rm V}^3\equiv \frac{GM}{\Lambda^2}
\end{equation}
is the \textit{Vainshtein Radius}
and $\Lambda$ is a new mass scale that appears in the underlying theory. Far outside the Vainshtein radius, we have $x\approx2\beta^2$ and so 
$F_5=2\beta^2 F_{\rm N}$, the fifth-force is a factor of $2\beta^2$ larger than the Newtonian one. This is the unscreened regime. When $r\ll r_{\rm 
V}$ we instead have
\begin{equation}
\frac{F_5}{F_{\rm N}}=2\beta^2\left(\frac{r}{r_{\rm V}}\right)^{\frac{3}{2}},
\end{equation}
and the fifth-force is suppressed by a factor of $( {r}/{r_{\rm V}})^{3/2}$. This is the screened regime. The Vainshtein radius of the Sun is 
$\mathcal{O}(\textrm{pc})$ \cite{Khoury:2013tda} and so one does not need to tune $\beta\ll1$ in order to satisfy solar system tests. 

\subsection{Breaking of the Vainshtein Mechanism and Stellar Structure}

The cubic galileon is the simplest example of a theory that exhibits the Vainshtein mechanism. \cite{Kobayashi:2014ida} have shown that the mechanism 
is partially broken inside objects of finite extent in more generic theories. Specifically, outside objects the mechanism works as above but their 
internal structure is governed by the modified hydrostatic equilibrium equation \cite{Koyama:2015oma,Saito:2015fza}
\begin{equation}\label{eq:MGHSE}
\frac{\dd P}{\dd r}=-\frac{GM\rho}{r^2}-\frac{\Upsilon}{4}G\rho\frac{\dd^2 M}{\dd r^2}.
\end{equation}
Here, $\Upsilon$ is a dimensionless parameter that characterises the strength of the modifications of gravity; it is a combination 
of the new mass scales appearing in the Lagrangian for the theory and the time-derivative of the cosmological field. Note that since the mass is 
more concentrated towards the centre of stars, the new term corresponds to a weakening of gravity if $\Upsilon>0$ and a strengthening when the 
converse is true.

In terms of the parameters 
appearing in the effective field theory of dark energy (EFT) \cite{Bellini:2014fua,Gleyzes:2014qga}, one has\footnote{In general, there are two 
additional parameters, $\alpha_K$ and $M_*$.} \cite{Saito:2015fza}
\begin{equation}
 \frac{\alpha_H^2}{\alpha_H-\alpha_T-\alpha_B(1+\alpha_T)}\lsim 0.35.
\end{equation}
The EFT describes the cosmology of beyond Horndeski theories, a very general scalar-tensor theory\footnote{Note that the beyond Horndeski class 
encompasses a wide variety of healthy scalar-tensor extensions of GR but it is by no means the most general 
\cite{Zumalacarregui:2013pma,Gao:2014soa,Gao:2014fra} theory.} where the equations of motion are second-order \cite{Gleyzes:2014qga}, on linear 
scales. The parameters therefore govern deviations from GR on these scales and upcoming surveys aimed at testing 
gravity will focus on them. Any independent constraint from small scales is therefore complementary to these searches. 
Furthermore, several viable competitors to $\Lambda$CDM, for example, the covariant quartic galileon, which has $\Upsilon=1/3$, predict 
$\Upsilon\sim\mathcal{O}(1)$ precisely because the dynamics of the theory differ greatly from GR. The constraint we will obtain from the MMHB is 
strong enough to rule these out. Note that $\Upsilon\propto\alpha_H^2$, which is only non-zero when the beyond Horndeski terms are present. This 
reflects the fact that the Vainshtein mechanism works flawlessly when the theory is pure Horndeski and so only those theories that contain some 
beyond Horndeski terms are probed. Since there are no symmetries protecting the general action, one would expect quantum corrections to generate such 
terms in generic modified gravity theories. Finally, we note that there is no upper limit on $\Upsilon$, but \cite{Saito:2015fza} have shown that 
theories where $\Upsilon<-2/3$ do not give stable stellar configurations. For this reason, we restrict our attention to the region 
$-2/3<\Upsilon<\infty$.

A wide variety of stellar systems can be described using a \textit{polytropic} equation of state (EOS)
\begin{equation}
P=K\rho^{\frac{n+1}{n}},
\end{equation}
where $K$ is a constant that depends on the composition of the fluid and $n$ is known as the \textit{polytropic index}. In GR, the equations of 
stellar structure are scale-invariant and this allows one to reformulate them in terms of dimensionless quantities. This symmetry is preserved by 
equation (\ref{eq:MGHSE}) and so we write $r=r\ccc\xi$, $\rho=\rho\ccc\theta(\xi)^n$ 
and $P=P\ccc\theta(\xi)^{n+1}$, where $P\ccc$ and $\rho\ccc$ are the central pressures and densities and
\begin{equation}\label{eq:rc}
 r\ccc^2\equiv\frac{(n+1)P\ccc}{4\pi \GN \rho\ccc^2}.
\end{equation}
Using the continuity equation, one has
\begin{equation}\label{eq:M'}
 \frac{\dd M}{\dd r}=4\pi r^2 \rho(r),
\end{equation}
which implies
\begin{equation}\label{eq:M''}
 \frac{\dd^2 M}{\dd r^2}=8\pi r \rho+4\pi r^2 \frac{\dd \rho}{\dd r}.
\end{equation}
Using these in conjunction with equation (\ref{eq:MGHSE}), one is led to the modified Lane-Emden equation \cite{Koyama:2015oma}
\begin{equation}\label{eq:MLE}
 \frac{1}{\xi^2}\frac{\dd}{\dd \xi}\left[\left(1+{\frac{n}{4}\Upsilon\xi^2\theta^{n-1}}\right)\xi^2\frac{\dd \theta}{\dd 
\xi}+\frac{\Upsilon}{2}\xi^3\theta^n\right]=-\theta^n,
\end{equation}
which is subject to the boundary conditions $\theta(0)=1$ ($P(0)=P\ccc$, $\rho(0)=\rho\ccc$) and $\theta'(0)=0$ ($\dd P/\dd r(0)=0$\footnote{This is 
a consequence of spherical symmetry.}). The radius $R$ of the star is defined by $P(R)=0$, which defines $\xi_R$ such that $\theta(\xi_R)=0$.

In GR, $\Upsilon=0$ and (\ref{eq:MLE}) reduces to the Lane-Emden equation. In this case, each polytropic index $n$ has a unique solution. In 
alternative theories, this is no longer the case and the solution varies with $\Upsilon$. It was shown in \cite{Koyama:2015oma} that the solution 
near 
the origin is given by
\begin{equation}\label{eq:thexpo}
 \theta(\xi)=1-\alpha\xi^2,\quad\textrm{with}\quad \alpha = \frac{1}{6}+\frac{\Upsilon}{4}.
\end{equation}

Three important dimensionless quantities are
\begin{align}
\omega_n&\equiv -\xi^2\left.\frac{\dd\theta}{\dd\xi}\right\vert_{\xi={\xi_R}},\label{eq:ome}\\
\gamma_n&\equiv(4\pi)^{\frac{1}{n-3}}(n+1)^{\frac{n}{3-n}}\omega_n^{\frac{n-1}{3-n}}\xi_R,\quad \textrm{and}\label{eq:gammadef}\\
\delta_n&\equiv-\frac{\xi_R}{3\dd\theta/\dd\xi|_{\xi=\xi_R}},
\end{align}
which appear in the formula for the mass
\begin{equation}\label{eq:MLE2}
M=4\pi r\ccc^3\rho\ccc\omega_n,
\end{equation}
the mass-radius relation
\begin{equation}\label{eq:MRrel}
R=\gamma_n\left(\frac{K}{G}\right)^{\frac{n}{3-n}}M^{\frac{n-1}{n-3}},
\end{equation}
and the central density
\begin{equation}
\rho\ccc=\delta_n\left[\frac{3M}{4\pi R^3}\right].\label{eq:rhoc}
\end{equation}

\section{Dwarf Stars}\label{sec:BD}

In this section, we provide a brief overview of the salient features of brown dwarf theory necessary for the calculations presented later on. The 
reader is referred to \cite{Burrows:1992fg} and references therein for a more detailed account.

The brown dwarf branch occupies the region between Jupiter-like gas planets ($M_{\rm J}=10^{-3}M_\odot$) and main-sequence stars 
($M\gsim0.08M_\odot$). They are composed of molecular hydrogen and helium in the liquid metallic phase, except in a very thin layer 
near the surface where the density is too low and the fluid exists as a weakly coupled plasma that satisfies the ideal gas law. They are fully 
convective and contract along the Hayashi track and therefore have a polytropic EOS with $n=1.5$ \cite{kippenhahn1990stellar}. High-mass brown dwarfs 
are supported by electron degeneracy pressure, 
but classical Coulomb corrections shift the EOS of low mass stars ($M\lsim 4 M_{\rm J}$) towards $n=1$ 
\cite{1996ApJ...460..993S,lissauer2013fundamental}. Equation 
(\ref{eq:MRrel}) reveals that, when this is the case, the radius 
is independent of the mass and is fixed by the EOS and the solution of the Lane-Emden equation alone. In GR, one finds\footnote{Note that the 
Lane-Emden equation has the exact solution $\theta(\xi)=\sinc(\xi)$ when $\Upsilon=0$ (i.e. GR) \cite{chandrasekhar2012introduction}, whose first 
zero occurs when $\xi=\pi$. Putting this into equation \eqref{eq:gammadef} gives $\gamma_1=\sqrt{\pi/2}$.} 
$\gamma_1=\sqrt{\pi/2}\approx 1.253$ and $R\approx 0.1R_\odot$. We will return to calculate this for alternative theories later in section 
\ref{sec:MGBD}.

When the star is first formed, it contracts under its own self-gravity, which leads to a rise in its central temperature and density. The contraction 
stops at the onset of either electron degeneracy, or thermonuclear fusion depending on the mass; this defines brown dwarf and red dwarf stars 
respectively. Only objects that are sufficiently heavy can reach 
central conditions capable of thermonuclear ignition before the fluid becomes degenerate. Thus, there 
is a minimum mass for hydrogen burning, which is typically around $0.08M_\odot$ if the theory is GR. We will calculate this mass in alternative 
theories in section \ref{sec:MGBD}. If the star cannot achieve thermonuclear ignition before degeneracy sets in, it will become a brown dwarf. 
Degenerate gasses 
have equations of state where the pressure is almost independent of the temperature and so radiation from the surface does not lead to further 
contraction. Instead, the brown dwarf cools over time \cite{1962iss..rept....1K,1963PThPh..30..460H,Burgasser:2009ym}.

Typically, the central temperature and density of brown dwarfs are of order $10^6$ K and $10^3$ g/cm$^3$ respectively. These conditions are not 
sufficient to 
overcome the Coulomb barrier for  $^3$He--$^3$He and $^3$He--$^4$He reactions and so burning cannot proceed via the PP chains. Instead, the star 
consumes deuterium in the following reactions:
\begin{align*}
\textrm{p}+\textrm{p}&\rightarrow\textrm{d}+\textrm{e}^++\nu_e\\
\textrm{p}+\textrm{d}&\rightarrow\,^3\!\!\,\,\textrm{He}+\gamma.
\end{align*}
The second reaction is a strong interaction and therefore proceeds quickly, consuming primordial deuterium. The first is a weak process creating a 
new source of deuterium and is therefore the rate-limiting, \textit{bottle-neck}, reaction. The energy 
generation rate per unit mass at brown dwarf temperatures and densities follows the approximate power-law form
\begin{equation}\label{eq:engen}
\epsilon_{\rm pp}=\epsilon\ccc\left(\frac{T}{T\ccc}\right)^s\left(\frac{\rho}{\rho\ccc}\right)^{u-1},
\end{equation}
with $s\approx 6.31$ and $u\approx 2.28$. $\epsilon\ccc$ is given by (see \cite{1973ApJ...181..457G,Fowler:1975kz,Burrows:1992fg})
\begin{equation}\label{eq:epsc}
\epsilon_c=\epsilon_0 X^2 T\ccc^s\rho\ccc^{u-1}\exp\left(\frac{0.147}{\mu\eee^{0.43}}\right),
\end{equation}
where $X$ is the hydrogen mass fraction and $\epsilon_0=5.2\times10^{-9}$ ergs/g/s. 

\section{Dwarf Stars in Alternative Gravity Theories}\label{sec:MGBD}

This section constitutes the main original results of this paper. The derivations will closely follow the semi-analytic model of Burrows \& Liebert 
\cite{Burrows:1992fg}.

\subsection{Brown Dwarfs: The Radius Plateau}

\begin{figure}
\includegraphics[width=0.45\textwidth]{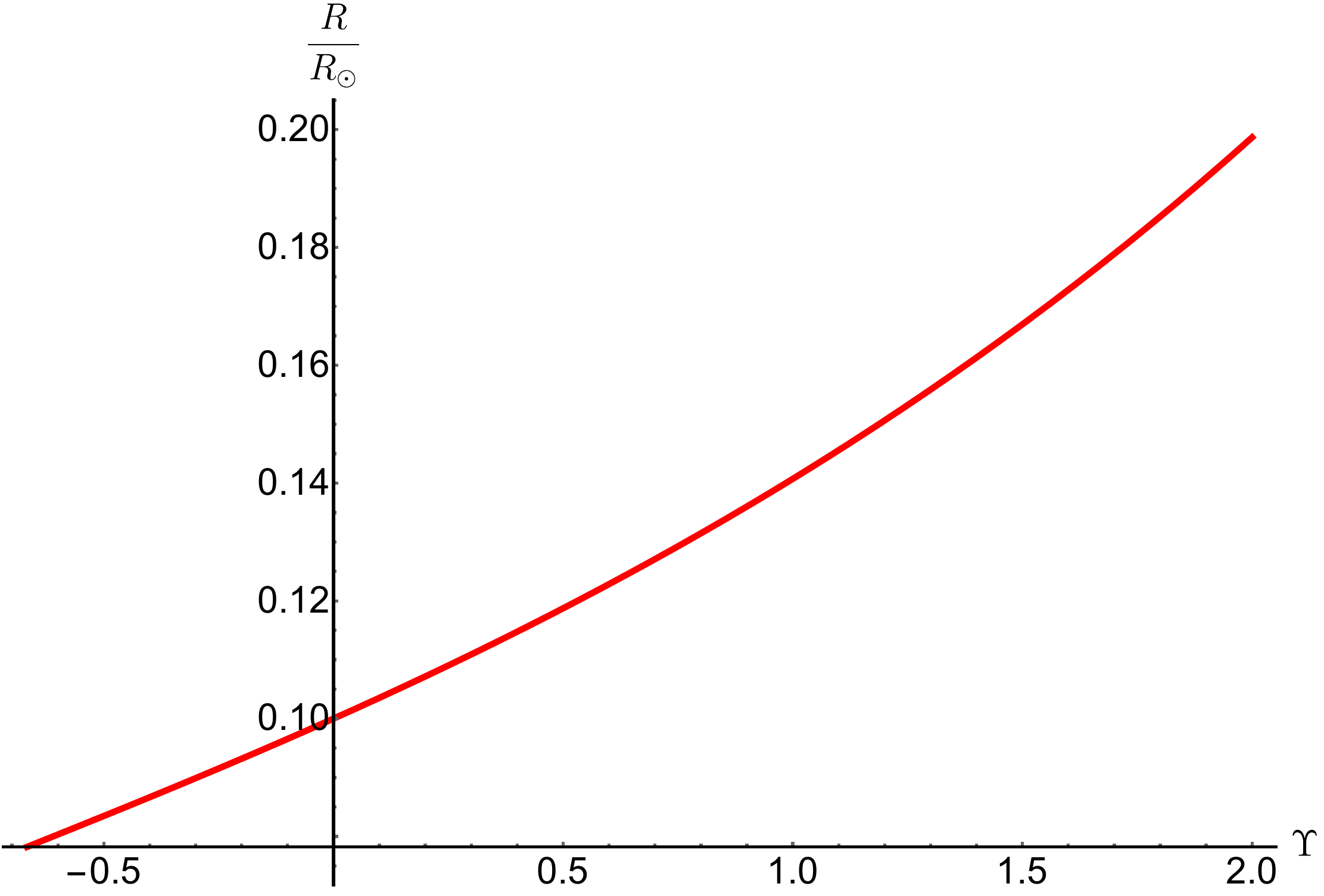}
\caption{The radius of low mass brown dwarf stars in alternative theories of gravity as a function of $\Upsilon$.}\label{fig:rad}
\end{figure}

As remarked in section \ref{sec:BD}, low mass brown dwarfs ($M\lsim 4M_{\rm J}$) are well-modelled by $n=1$ polytropes and hence have radii given by
\begin{equation}
R=\gamma_1\left( \frac{K}{G}\right)^{\frac{1}{2}},
\end{equation}
which is $\approx 0.1 R_\odot$ when $\Upsilon=0$ i.e. in GR\footnote{The constant $K\approx2.7\times10^5$ m$^5$/kg/s$^2$, which is a fit to a 
more precise EOS \cite{1996ApJ...460..993S,lissauer2013fundamental}.}. Since $K$ is independent of the theory of gravity, the radius of 
low mass brown dwarfs 
in alternative theories of gravity is given by
\begin{equation}
R(\Upsilon)=0.1\frac{\gamma_1(\Upsilon)}{\gamma_1(\Upsilon=0)}R_\odot.
\end{equation}
This is plotted as a function of $\Upsilon$ in figure \ref{fig:rad}. When $\Upsilon>0$, gravity is weaker and the radius is larger than the GR 
prediction. When $\Upsilon<0$, gravity is stronger and hence the radius is lower, up to the limiting value of $\Upsilon=-2/3$, which predicts 
$R=0.078 R_\odot$.
%
\cite{Koyama:2015oma} have investigated the changes in main-sequence and 
red giant branch stars these theories predict and have found that they are $\lsim $1\% when $\Upsilon\lsim \mathcal{O}(1)$. This is the first 
example showing that the changes to the structure of brown dwarfs can still be significant in this range and that these stars are better probes of 
alternative theories of gravity.  Unfortunately, measurements of the mass-radius relation for low mass objects are 
sparse (see \cite{Chabrier:2008bc,2041-8205-810-2-L25}, figure 2) and so we will not attempt to constrain $\Upsilon$ using this effect. Future data releases from 
surveys targeted at low mass objects such as GAIA \cite{Sozzetti:2003vn,2014MNRAS.437..497S} will improve the empirically measured mass-radius 
relation and therefore have the potential to place new constraints on alternative gravity theories.


\subsection{High Mass Brown Dwarfs: The Minimum Mass for Hydrogen Burning}

Higher mass brown dwarfs are supported by degeneracy pressure. A good measure of the degree of degeneracy is 
\begin{equation}\label{eq:etadef}
\eta\equiv \frac{\mu_{\rm F}}{k_{\rm B}T}=\frac{(3\pi^2)^{\frac{2}{3}}\hbar^2}{2m_{\rm e}m_{\rm H}^{\frac{2}{3}}k_{\rm 
B}}\frac{\rho^{\frac{2}{3}}}{\mu_{\rm e}^{\frac{2}{3}}T},
\end{equation} 
where $\mu_{\rm F}$ is the Fermi energy, $m_{\rm H}$ is the mass of atomic hydrogen, and $\mu_{\rm e}$ is the number of baryons per electron. Note 
that $\rho^{1/n}/T$ is a constant for polytropic gasses \cite{horedt2006polytropes} and so $\eta$ is a constant. For a metal poor gas with $Z=0$ (i.e. 
zero metallicity), which is appropriate for brown dwarfs, one has
\begin{equation}
\frac{1}{\mu_{\rm e}}=X+\frac{Y}{2}=\frac{X+1}{2},
 \end{equation}
with $X$ and $Y$ being the hydrogen and helium mass fractions respectively. In this work we will not fix $\mu_{\rm e}$ in order to show the 
dependency of important quantities, however we note that for a hydrogen-helium fluid with $X=0.75$, $Y=0.25$, which is typical for a brown dwarf, one 
finds $\mu_{\rm e}=1.143$, which we will take when we compute any numerical values.
 
When the gas is fully degenerate, the pressure can be found by integrating over the Fermi-Dirac distribution to find 
\begin{equation}
P_{\rm deg}=\frac{(3\pi^2)^{\frac{2}{3}}\hbar^2}{5m\eee m_{\rm H}^{\frac{5}{3}}}\frac{\rho^{\frac{5}{3}}}{\mu\eee^{\frac{5}{3}}}.
\end{equation}
In the other extreme, when the pressure support is due to the internal motion of the gas, the EOS is given by the idea gas law
\begin{equation}
P_{\rm gas}=\frac{\rho k_{\rm B}T}{\mu m_{\rm H}},
\end{equation} 
 where $\mu$ is the mean molecular mass given by
 \begin{equation}
 \frac{1}{\mu}=2X+\frac{3Y}{4}.
 \end{equation}
Again, we will not fix the value of $\mu$ in order to show the dependency of intermediate expressions, but we note that $\mu=0.593$ for $X=0.75$, 
$Y=0.25$. We will use this value when evaluating any expressions numerically. Using equation 
(\ref{eq:etadef}), one has $P_{\rm gas}=\alpha P_{\rm deg}/\eta$, with
\begin{equation}
\alpha\equiv \frac{5\mu\eee}{2\mu}\approx 4.82.
\end{equation}
In general, a good approximation for the EOS between the two regimes is
\begin{equation}
P=\frac{(3\pi^2)^{\frac{2}{3}}\hbar^2}{5m\eee m_{\rm 
H}^{\frac{5}{3}}}\left(1+\frac{\alpha}{\eta}\right)\frac{\rho^{\frac{5}{3}}}{\mu\eee^{\frac{5}{3}}},
\end{equation} 
 and so one can see that the equation of state is polytropic with index $n=3/2$ and polytropic constant
 \begin{equation}\label{eq:Kfund} K=\frac{(3\pi^2)^{\frac{2}{3}}\hbar^2}{5m\eee m_{\rm 
H}^{\frac{5}{3}}\mu\eee^{\frac{5}{3}}}\left(1+\frac{\alpha}{\eta}\right).
 \end{equation}
 
Equation (\ref{eq:Kfund}) can then be used in equation (\ref{eq:MRrel}) to find the stellar radius
\begin{equation}
R=\frac{(3\pi^2)^{\frac{2}{3}}}{5G m\eee m_{\rm H}^\frac{5}{3}\mu\eee^{\frac{5}{3}}}\gamma_{3/2}M^{-\frac{1}{3}}\left(1+\frac{\alpha}{\eta}\right),
\end{equation}
which can be substituted into equation (\ref{eq:rhoc}) to find the core density
\begin{equation}
\rho\ccc=\frac{125 G^3 m\eee^3  m_{\rm 
H}^5\mu\eee^5}{12\pi^5\hbar^6}\frac{\delta_{3/2}}{\gamma_{3/2}^3}M^2\left(1+\frac{\alpha}{\eta}\right)^{-3}.\label{eq:rhocfull}
\end{equation} 
The core temperature can be found using equation (\ref{eq:etadef}):
\begin{equation}
T\ccc=\frac{25 G^2 m\eee m_{\rm H}^{\frac{8}{3}} \mu\eee^{\frac{8}{3}}}{2^{\frac{7}{3}}\pi^2 k_{\rm 
B}\hbar^2}\frac{\delta_{3/2}^{\frac{2}{3}}}{\gamma_{3/2}^2}\frac{\eta}{(\alpha+\eta)^2}M^{\frac{4}{3}}\label{eq:Tcfull}.
\end{equation}
In order to find the luminosity $L_{\rm HB}$ from hydrogen burning, we must integrate the energy generation rate (\ref{eq:engen}) over the volume of 
the star. Since $\eta$ is constant we have $T/T\ccc=(\rho/\rho\ccc)^{\frac{2}{3}}$ from equation (\ref{eq:etadef}), which implies 
\begin{equation}\label{eq:lnint1}
L_{\rm HB}=4\pi r\ccc^3\rho\ccc\epsilon\ccc\int_0^{\xi_R}\xi^2\theta^{n(u+\frac{2}{3}s)}\dd\xi.
\end{equation}
Using equation (\ref{eq:thexpo}), we have 
\begin{equation}
\theta(\xi)\approx 1-(1+\frac{3\Upsilon}{2})\frac{\xi^2}{6}\approx \exp\left[-(1+\frac{3\Upsilon}{2}\frac{\xi^2}{6})\right].\label{eq:approxthet}
\end{equation}
One can then perform the integral (\ref{eq:lnint1}) and use equation (\ref{eq:MLE2}) to find\footnote{One may worry that the approximation 
\eqref{eq:approxthet} is not accurate since it extends the domain of integration to $\infty$ and is only valid to $\mathcal{O}(\xi^2)$. It is 
straightforward to verify that the difference between the integral and our approximation is always smaller than $10^{-2}$ over the range 
$0\le\Upsilon\le0.05$, which is the range we will investigate below. Physically, the energy generated through nuclear burning is dominated by the 
central region of the star since the density and temperature fall off quickly and so the contribution to $L_{\rm HB}$ comes mainly from the region 
$\xi\ll1$.}
\begin{equation}
L_{\rm HB}=\frac{3\sqrt{3\pi}}{\sqrt{2}\omega_{3/2}\left[(1+\frac{3\Upsilon}{2})(\frac{3}{2}u+s)\right]^{\frac{3}{2}}}\epsilon\ccc M.
\end{equation}
Using $u=2.28$, $s=6.31$, equations (\ref{eq:epsc}), (\ref{eq:rhocfull}) and (\ref{eq:Tcfull}) and inserting the numerical values for the constants 
one finds
\begin{align}
\frac{{L_{\rm HB}}}{{5.2\times10^6 L_\odot} }= 
\frac{\delta_{3/2}^{5.487}}{\omega_{3/2}\gamma_{3/2}^{16.46}(1+\frac{3\Upsilon}{2})^{\frac{3}{2}}}M_{-1}^{11.973}\frac{\eta^{10.15}}{
(\eta+\alpha)^{16.46}},
\end{align}
where $M_{-1}=M/0.1M_\odot$.
 
Our next task is to compute the luminosity at the photosphere\footnote{We denote photosphere quantities with the subscript $\textrm{e}$.}. This is 
the radius where the optical depth, defined as
\begin{equation}\label{eq:optdepth}
\tau(r)=\int_r^\infty \kappa_{\rm R} \rho\dd r,
\end{equation}
where $\kappa_{\rm R}$ the \textit{Rosseland mean opacity}, is equal to $2/3$. It is here that observations of quantities such as the luminosity and 
temperature probe and not the stellar surface. The photosphere typically lies very close to the stellar radius and so the fluid is in the gas 
phase rather than the metallic one. In GR, one finds the pressure at the optical depth by treating the surface gravity 
\begin{equation}
g\equiv\frac{GM(r)}{r^2}\approx\frac{GM(R)}{R^2}
\end{equation}
as a constant throughout the photosphere and integrating the hydrostatic equilibrium equation\footnote{This is justified because the distance from 
the photosphere to the surface is small compared with the stellar radius.}. In alternative theories, we have the additional complication coming from 
the 
additional term in equation (\ref{eq:MGHSE}). We will deal with this using approximations consistent with those made in GR. In particular, setting 
$g$ 
to be constant implies $\dd (M/r^2)/\dd r=0$ and so we have
 \begin{equation}
 \frac{\dd M}{\dd r}=2\frac{M}{r},
 \end{equation}
 which can be differentiated again to find
 \begin{equation}
 \frac{\dd ^2 M}{\dd r^2}=2\frac{M}{r^2}.
  \end{equation}
  Equation (\ref{eq:MGHSE}) is then
  \begin{equation}
  \frac{\dd P\eee}{\dd r}=-g\rho\left(1+\frac{\Upsilon}{2}\right),
  \end{equation}
  which can be integrated from the photosphere to find
\begin{equation}\label{eq:Pphot}
  P\eee=\frac{2}{3\kar}\left(1+\frac{\Upsilon}{2}\right)g,
\end{equation}
where equation (\ref{eq:optdepth}) has been used. 
Note that modified gravity is then partly degenerate with reducing the opacity\footnote{We use the term \textit{partly} because there are other places 
where modified gravity 
has effects that cannot be compensated by changing the opacity.}. Indeed, one can define an effective opacity
\begin{equation}\label{eq:effop}
 \kar^{\rm eff}=\frac{\kar}{\left(1+\frac{\Upsilon}{2}\right)},
\end{equation}   
which brings all of the equations describing the properties of the photosphere into the same form as GR.
Using the ideal gas law, equation (\ref{eq:Pphot}) becomes
\begin{equation}\label{eq:preseq}
\frac{\rho k_{\rm B} T}{\mu m_{\rm H}}=\frac{2}{3\kar}\left(1+\frac{\Upsilon}{2}\right)g.
\end{equation}
Unfortunately, this is all the analytic progress we can make with the stellar structure equations alone. In order to calculate the effective 
temperature, we need to know more about the metallic phase transition. To this, we use the approximate analytic fits to the specific entropy 
($s=Sm_{\rm H}/k_{\rm B}$) in the gas phase
\begin{equation}
s_{\rm gas}=-1.594\ln\eta+12.43,
\end{equation}
taken from \cite{1989ApJ...345..939B} and the metallic phase
\begin{equation}
s_{\rm metallic}=1.032\ln\left(\frac{T}{\rho^{0.42}}\right)-2.438,
\end{equation}
taken from \cite{Stevenson:1991eq}. Here, $T$ is measured in K and $\rho$ in g/cm$^3$. The photosphere is the radius where $s_{\rm gas}=s_{\rm 
metallic}$ and so we have 
\begin{equation}\label{eq:Trhoana}
T\eee=1.8\times10^6 \frac{\rho\eee^{0.42}}{\eta^{1.545}}\,{\textrm{K}}.
\end{equation}
Using equation (\ref{eq:MRrel}), the surface gravity is 
\begin{align}
g&=\frac{25G^3m\eee^2m_{\rm 
H}^{\frac{10}{3}}\mu\eee^{\frac{10}{3}}}{(81\pi^8)^{\frac{1}{3}}\hbar^4}\frac{M^{\frac{5}{3}}}{\gamma_{3/2}^2}\left(1+\frac{\alpha}{\eta}\right)^{-2}
\\&=\frac{3.15\times10^{6}}{\gamma_{3/2}^{2}}\left(\frac{M}{0.01M_\odot}\right)^{\frac{5}{3}}\left(1+\frac{ 
\alpha}{\eta}\right)^{-2}\textrm{cm}/\textrm{s}^2.
\end{align}
Using this and the polytropic relation (\ref{eq:Kfund}) in equation (\ref{eq:preseq}) we find
\begin{align}\label{eq:rhocana}
\frac{\rho\eee}{\textrm{g}/\textrm{cm}^3}=5\times10^{-5}M_{-1}^{1.17} 
\left[\frac{(1+\frac{\Upsilon}{2})}{\kappa_{-2}}\right]^{0.7}\frac{\eta^{1.09}}{\gamma_{3/2}^{1.41}}\left(1+\frac{\alpha}{\eta
}\right)^{-1.41},
\end{align}
where $\kappa_{-2}=\kappa/10^{-2}\,\,\textrm{cm}^2/\textrm{g}$.
Using this in equation (\ref{eq:Trhoana}) we find the effective temperature
\begin{equation}\label{eq:Teff}
\frac{T\eee}{\textrm{K}}=2.9\times10^4\frac{ M_{-1}^{0.49} }{\gamma_{3/2}^{0.59}\eta^{1.09}}
\left[\frac{(1+\frac{\Upsilon}{2})}{\kar}\right]^{0.296}\left(1+\frac{\alpha}{\eta}
\right)^{-0.59}.
\end{equation}
The stellar luminosity $L\eee$ is then found by inserting equations ({\ref{eq:MRrel}) and (\ref{eq:Teff}) into the formula $L\eee=4\pi R^2\sigma 
T\eee^4$ to find
\begin{equation}\label{eq:L}
L\eee=2.65L_\odot 
\frac{M_{-1}^{1.305}}{\gamma_{3/2}^{2.366}\eta^{4.351}}\left[\frac{(1+\frac{\Upsilon}{2})}{\kar}\right]^{1.183}\left(1+\frac{\alpha}{\eta}\right)^{
-0.366}.
\end{equation} 

The star can burn hydrogen stably when $L_{\rm HB}=L\eee$ and so equating (\ref{eq:lnint1}) and (\ref{eq:L}) one has
\begin{equation}\label{eq:etafunc}
3.76 M_{-1}
=\left[\frac{\left(1+\frac{\Upsilon}{2}\right)}{\kar}\right]^{0.11}\left(1+\frac{3\Upsilon}{2} 
\right)^{0.14}\frac{\gamma_{3/2}^{1.32}\omega_{3/2}^{0.09}}{\delta_{3/2}^{0.51}}I(\eta), 
\end{equation}
with
\begin{equation}
I(\eta)\equiv \frac{(\alpha+\eta)^{1.509}}{\eta^{1.325}}.
\end{equation}
From here on we set $\kar=10^{-2}$ cm$^2$/g, which is typical for high-mass brown dwarfs. The stellar composition does not vary between different 
stars by large amounts and since equation (\ref{eq:etafunc}) is a weak function of the opacity, the results we will obtain are largely 
independent of this choice. The weak opacity-dependence is one of the reasons dwarf stars are excellent probes of modified gravity. Importantly, the 
function $I(\eta)$ has a minimum value of $2.34$ when $\eta=34.7$. This means that if $M$ is too low, there is no consistent solution to equation 
(\ref{eq:etafunc}) and hence there is a minimum mass for hydrogen burning. In GR, one has $\gamma_{3/2}=2.357$, $\delta_{3/2}=5.991$ and 
$\omega_{3/2}=2.714$, which gives $M_{\rm MMHB}^{\rm GR}\approx 0.08M_\odot$. Remarkably, this simple analytic model makes a prediction that is very 
close to the results of more detailed simulations \cite{1963ApJ...137.1121K,Burrows:1997ka}, which predict a value of $0.075M_\odot$. In figure 
\ref{fig:mass} we plot 
the MMHB predicted by our alternative theory of gravity as a function of $\Upsilon$; one can see that the MMHB is very sensitive to its value. Large 
values of $\Upsilon>0$\footnote{As discussed above, we will not consider the case $\Upsilon<0$ here because this lowers the MMHB and is hence always 
consistent with observations of low mass M-dwarfs.} raise the MMHB above the GR prediction. This is because the reduced gravity ensures that stars of 
fixed mass have cores that are cooler and less dense because less nuclear burning is needed to provide the pressure gradient that prevents 
gravitational collapse. Note that, according to equation \eqref{eq:L}, there is a small effect on the photospheric properties due to modified 
gravity. These are largely sub-dominant because the are degenerate with changing the opacity, which effects the MMHB as $\kar^{0.11}$. Most of the 
deviations from GR are due to changes in the core conditions. Note from equation \eqref{eq:engen} that the rate of hydrogen burning is a strong 
function of the core temperature and pressure, even stronger than the rate for PP chain burning in main-sequence stars\footnote{The energy 
generation on the PP chains is $\epsilon_{\rm PP}\propto \rho T^4$.}. This is one reason why dwarf stars are more sensitive probes of alternative 
gravity theories.

There have been several observations of low mass Red Dwarf stars in the Milky Way with masses in the range 
$0.08$--$0.1M_\odot$ \cite{Coppenbarger:1994us,1996A&A...315..418B,Delfosse:2000jr,Martinache:2006ud,hamilton2012our}. Clearly, large values of 
$\Upsilon$ are in tension with these observations since they predict that these objects would be brown dwarfs. The lowest mass observed M-dwarf is Gl 
866 C \cite{Segransan:2000jq}; it's measured mass is $0.0930\pm0.0008\,M_\odot$. Using equation 
(\ref{eq:etafunc}), consistency with this measurement is only achieved when $\Upsilon\lsim1.6$; values larger than this are excluded. In 
particular, this places a new constraint on the parameters appearing in the EFT:
\begin{equation}\label{eq:cons}
 \frac{\alpha_H^2}{\alpha_H-\alpha_T-\alpha_B(1+\alpha_T)}\lsim 0.41.
\end{equation}
%

We end this section by discussing the possible astrophysical degeneracies that could mimic the effects of modified gravity. These are few. We have 
seen 
already that the theories considered here are partly degenerate with changing the opacity (see equation (\ref{eq:effop})). Since the MMHB is only a 
weak function of the opacity ($\propto \kar^{-0.11}$), any variation has only a minimal effect and cannot mimic the large changes induced by 
modified gravity. The MMHB is also weakly dependent on the amount of stellar rotation, which was absent in our model. Rotation acts to increase the 
MMHB \cite{1992ApJ...393..258S,1970A&A.....8...50K} and so can only compound the effect presented here. In order to make a firm prediction, we 
had to 
assume specific values for the composition parameters $\mu\eee$ and $\mu$. Brown dwarfs show little chemical evolution over their 
lifetime (partly because their lifetime is so long) and so these are always $\mathcal{O}(1)$. Therefore, our prediction is robust to variations in 
their values. There are also the effects of missing physics. Our model ignored the effects of mixing and other transport 
processes. We also assumed an EOS in order to make the problem analytically tractable. When these effects are included in full numerical 
models \cite{1963ApJ...137.1121K}, one finds that the MMHB changes by less than $0.01M_\odot$, which is far smaller than the changes due to modified 
gravity. The missing physics is non-gravitational and so it is reasonable to expect similar discrepancies if one were to use full numerical models 
including the change to the hydrostatic equilibrium equation. The potential changes are an order-of-magnitude smaller than the changes due to 
modified gravity and so the conclusions here should remain largely unchanged, although one should confirm this numerically. 

Finally, one must be certain that the empirical mass determination of low mass stars does not implicitly assume GR. If this is the case, our 
constraint is not self-consistent. Fortunately, this is not the case. The theories considered here do not exhibit deviations from GR outside 
astrophysical bodies and Newtonian physics governs the motion of binary objects. The masses of some of the above referenced stars are determined 
using either the eclipsing binary technique, or by measuring the motions of their satellites. 
These methods both rely on Newtonian mechanics and not the object's intrinsic properties. For this reason, the mass determination is independent 
of whether the theory is GR or scalar-tensor. If the stars are not in binaries, or do not have any observable satellites, photometry can used to 
measure 
the mass using the mass-luminosity relation. One can see that equation (\ref{eq:L}) implies that this is sensitive the theory of 
gravity. The relation used to infer the photometric mass is empirical and not theoretically determined. Indeed, it is calibrated using observations 
of low mass stars with known mass found using the eclipsing binary technique \cite{1993AJ....106..773H}. For this reason, masses found using this 
method are insensitive to the theory of gravity.

\begin{figure}
\includegraphics[width=0.45\textwidth]{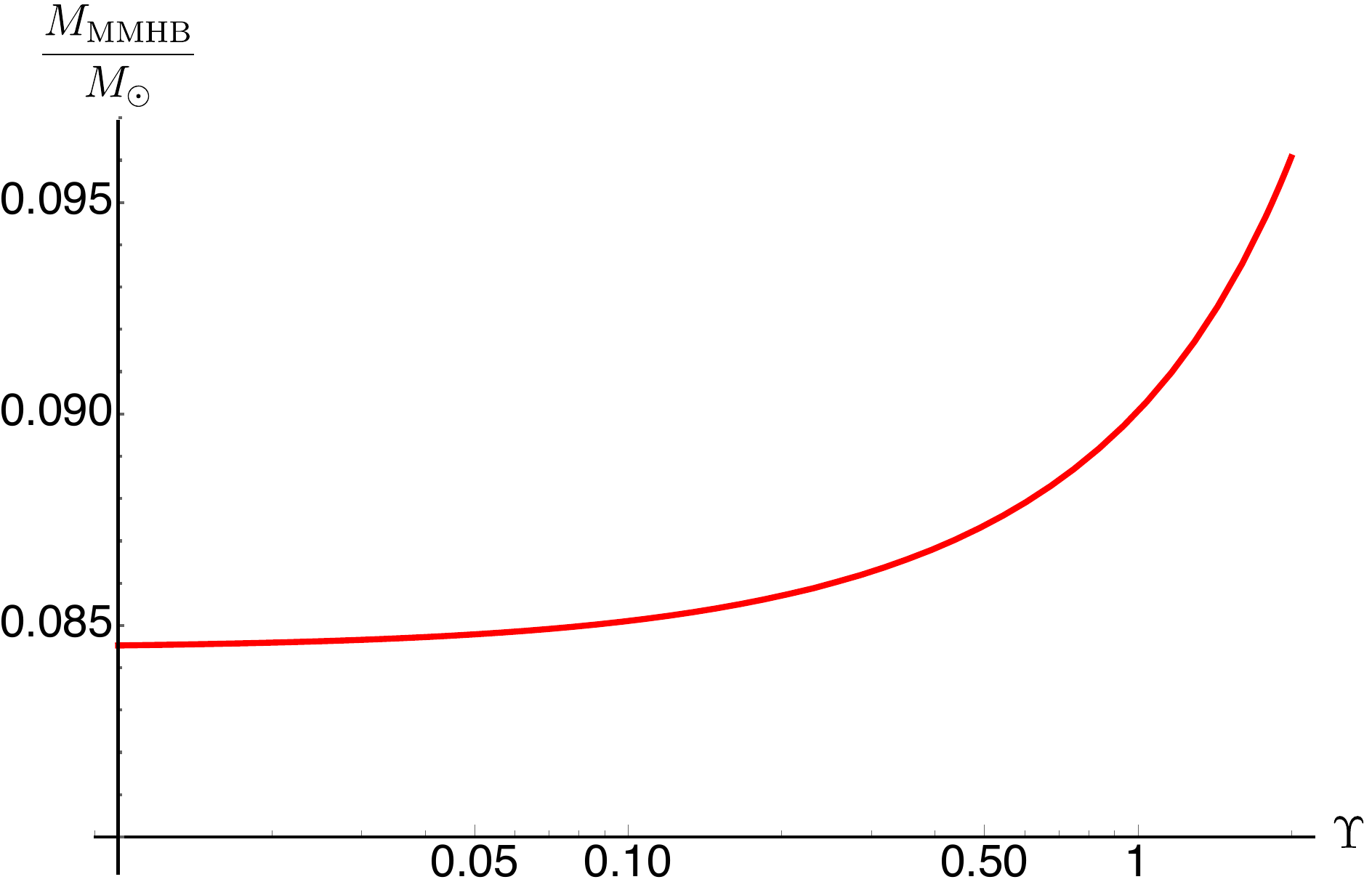}
\caption{The MMHB as a function of $\Upsilon$.}\label{fig:mass}
\end{figure}

\section{Conclusions}\label{sec:concs}

In this paper, we have pointed out that low mass stellar objects, in particular brown and red dwarf stars, are excellent probes of a new an 
interesting 
class of alternative theories of gravity: the beyond Horndeski class. Unlike main- and post-main-sequence stars, there are few astrophysical 
degeneracies and these are weak functions of non-gravitational physics. Furthermore, we have seen that deviations from the GR 
predictions are non-negligible over a larger range of $\Upsilon$ than in objects such as main-sequence stars and dark matter haloes. For example, 
main-sequence stars show negligible deviations when $\Upsilon\lsim\mathcal{O}(1)$; this is not the case for dwarf stars.

Low-mass brown dwarfs have radii that are largely independent of their mass. Here, we have presented a simple analytic model 
of these objects based on $n=1$ polytropes, which make a unique prediction for the radius in terms of fundamental constants and the solution of the 
Lane-Emden equation. We found that this is smaller than the GR prediction of $0.1R_\odot$ when $\Upsilon<0$ (stronger gravity) and can be as small 
as $R=0.078 R_\odot$, corresponding to the minimum value of $\Upsilon=-2/3$. When $\Upsilon>0$, the strength of gravity is reduced and the radius 
can be arbitrarily large. These are robust predictions of the theory and more realistic models predict that the physics missing in our 
model\footnote{The $n=1$ polytropic approximation is incomplete and a more realistic EOS is needed.} only cause small deviations in the 
radius. This means that the effects of modified gravity are not 
degenerate with the EOS. We do not claim any firm constraint here due to the sparse data in the mass range where our model applies (see 
\cite{Chabrier:2008bc} and references therein), but we note that future surveys such as GAIA, which will probe this mass range, could place new and 
independent constraints on $\Upsilon$ and the parameters appearing in the EFT. In particular, the parameter range $-2/3\le\Upsilon\le0$ is not 
probed by the MMHB, but makes the unique prediction that the radius plateau lies at radii $0.078<R/R_\odot<0.1$.

Turning to higher mass brown dwarfs and the brown-red dwarf transition, the minimum mass for hydrogen burning is an extremely sensitive probe 
of modified gravity. Hydrogen burning requires sufficiently high core temperatures and densities that only sufficiently heavy stars can achieve. 
General relativity predicts that the MMHB is $\approx 0.08M_\odot$. Here, we have incorporated the effects of alternative 
gravity theories into the semi-analytic model of \cite{Burrows:1992fg} to predict the MMHB as a function of $\Upsilon$. We have found that when 
$\Upsilon>0$, the MMHB is larger than the GR prediction due to the weakening of gravity requiring larger mass stars to achieve the necessary core 
conditions. Using the upper bound on the mass of the lightest observed M-dwarf, we were able to place the new constraint $\Upsilon\lsim1.6$, 
which translates into the bound \eqref{eq:cons} on the parameters appearing in the effective field theory of dark energy. It is interesting to note 
that the beyond Horndeski covariant quartic galileon model, which admits self-accelerating cosmological solutions and is therefore a competitor to 
$\Lambda$CDM, predicts $\Upsilon=1/3$ and is hence excluded by this constraint.

The parameters appearing in the EFT completely characterise the cosmology of beyond Horndeski theories, a very general class of 
healthy scalar-tensor extensions of GR, on linear scales. Any non-cosmological constraint restricts the possible deviations on these scales and is 
complementary to current and upcoming surveys that will probe the structure of gravity in this regime.

Finally, we end by discussing the applicability of the new tests we have presented here to other alternative gravity theories. The test required 
deviations from GR in the Newtonian limit and so cannot be applied to any theories that fit into the parametrised post-Newtonian framework, which 
predicts the inverse-square law. These include theories such as massless scalars and disformal theories 
\cite{EspositoFarese:2004cc,Sakstein:2014isa,Sakstein:2014nfa,Ip:2015qsa,Sakstein:2015jca}. Theories such as massive scalars and vectors as well as 
screened theories such as chameleons \cite{Khoury:2003rn} do predict deviations 
\cite{will1993theory,Hui:2009kc} and so, in principle, one could apply the tests here to 
these theories, although in the latter case a comparison with data would be incredibly difficult because one requires objects in other galaxies. Only 
theories that preserve the scale-invariant nature of the stellar structure equations will result in a modified Lane-Emden equation, theories that do 
not have this property, such as massive scalars, would require more advanced techniques to extract the MMHB and radius plateau. One important point 
to make is that the MMHB is very constraining for theories that predict a weakening of the gravitational force but not for theories that predict an 
enhancement. In this case, the MMHB is lower than the GR prediction and so the observation of M-dwarfs with masses compatible with GR do not disagree 
with these theories. Theories where gravity is weaker than the GR prediction are rare because they typically involve extra degrees of freedom such as 
scalars that couple directly to matter. Since the fifth-force is proportional to the square of this coupling, it is difficult to find repulsive 
forces. Our bound here was obtained using the MMHB but ultimately it may prove the case that the radius plateau is the more versatile of the tests we 
have presented here. With upcoming surveys providing better measurements of the mass-radius relation for low-mass objects, this new test of gravity 
provides another avenue to constrain other modifications of general relativity.

\begin{acknowledgments}
I am indebted to Olivier Asin for pointing out a typo in equation \eqref{eq:MLE} which led to figures \ref{fig:rad} and \ref{fig:mass} as well as the bound on $\Upsilon$ being incorrect in the published version of this work. I would like to thank Kazuya Koyama for many interesting discussions and a careful reading of the manuscript. I am grateful to Michael Cushing, Trent 
Dupuy, John Gizis, Artie Hatzes, Derek Homeier and Adam R. Solomon for several enlightening discussions.
\end{acknowledgments}
\bibliography{ref}

\begin{thebibliography}{62}%
\makeatletter
\providecommand \@ifxundefined [1]{%
 \@ifx{#1\undefined}
}%
\providecommand \@ifnum [1]{%
 \ifnum #1\expandafter \@firstoftwo
 \else \expandafter \@secondoftwo
 \fi
}%
\providecommand \@ifx [1]{%
 \ifx #1\expandafter \@firstoftwo
 \else \expandafter \@secondoftwo
 \fi
}%
\providecommand \natexlab [1]{#1}%
\providecommand \enquote  [1]{``#1''}%
\providecommand \bibnamefont  [1]{#1}%
\providecommand \bibfnamefont [1]{#1}%
\providecommand \citenamefont [1]{#1}%
\providecommand \href@noop [0]{\@secondoftwo}%
\providecommand \href [0]{\begingroup \@sanitize@url \@href}%
\providecommand \@href[1]{\@@startlink{#1}\@@href}%
\providecommand \@@href[1]{\endgroup#1\@@endlink}%
\providecommand \@sanitize@url [0]{\catcode `\\12\catcode `\$12\catcode
  `\&12\catcode `\#12\catcode `\^12\catcode `\_12\catcode `\%12\relax}%
\providecommand \@@startlink[1]{}%
\providecommand \@@endlink[0]{}%
\providecommand \url  [0]{\begingroup\@sanitize@url \@url }%
\providecommand \@url [1]{\endgroup\@href {#1}{\urlprefix }}%
\providecommand \urlprefix  [0]{URL }%
\providecommand \Eprint [0]{\href }%
\providecommand \doibase [0]{http://dx.doi.org/}%
\providecommand \selectlanguage [0]{\@gobble}%
\providecommand \bibinfo  [0]{\@secondoftwo}%
\providecommand \bibfield  [0]{\@secondoftwo}%
\providecommand \translation [1]{[#1]}%
\providecommand \BibitemOpen [0]{}%
\providecommand \bibitemStop [0]{}%
\providecommand \bibitemNoStop [0]{.\EOS\space}%
\providecommand \EOS [0]{\spacefactor3000\relax}%
\providecommand \BibitemShut  [1]{\csname bibitem#1\endcsname}%
\let\auto@bib@innerbib\@empty
\bibitem [{\citenamefont {Clifton}\ \emph {et~al.}(2012)\citenamefont
  {Clifton}, \citenamefont {Ferreira}, \citenamefont {Padilla},\ and\
  \citenamefont {Skordis}}]{Clifton:2011jh}%
  \BibitemOpen
  \bibfield  {author} {\bibinfo {author} {\bibfnamefont {T.}~\bibnamefont
  {Clifton}}, \bibinfo {author} {\bibfnamefont {P.~G.}\ \bibnamefont
  {Ferreira}}, \bibinfo {author} {\bibfnamefont {A.}~\bibnamefont {Padilla}}, \
  and\ \bibinfo {author} {\bibfnamefont {C.}~\bibnamefont {Skordis}},\ }\href
  {\doibase 10.1016/j.physrep.2012.01.001} {\bibfield  {journal} {\bibinfo
  {journal} {Phys.Rept.}\ }\textbf {\bibinfo {volume} {513}},\ \bibinfo {pages}
  {1} (\bibinfo {year} {2012})},\ \Eprint {http://arxiv.org/abs/1106.2476}
  {arXiv:1106.2476 [astro-ph.CO]} \BibitemShut {NoStop}%
\bibitem [{\citenamefont {Joyce}\ \emph {et~al.}(2014)\citenamefont {Joyce},
  \citenamefont {Jain}, \citenamefont {Khoury},\ and\ \citenamefont
  {Trodden}}]{Joyce:2014kja}%
  \BibitemOpen
  \bibfield  {author} {\bibinfo {author} {\bibfnamefont {A.}~\bibnamefont
  {Joyce}}, \bibinfo {author} {\bibfnamefont {B.}~\bibnamefont {Jain}},
  \bibinfo {author} {\bibfnamefont {J.}~\bibnamefont {Khoury}}, \ and\ \bibinfo
  {author} {\bibfnamefont {M.}~\bibnamefont {Trodden}},\ }\href@noop {} {\
  (\bibinfo {year} {2014})},\ \Eprint {http://arxiv.org/abs/1407.0059}
  {arXiv:1407.0059 [astro-ph.CO]} \BibitemShut {NoStop}%
\bibitem [{\citenamefont {Chang}\ and\ \citenamefont
  {Hui}(2010)}]{Chang:2010xh}%
  \BibitemOpen
  \bibfield  {author} {\bibinfo {author} {\bibfnamefont {P.}~\bibnamefont
  {Chang}}\ and\ \bibinfo {author} {\bibfnamefont {L.}~\bibnamefont {Hui}},\
  }\href@noop {} {\  (\bibinfo {year} {2010})},\ \Eprint
  {http://arxiv.org/abs/1011.4107} {arXiv:1011.4107 [astro-ph.CO]} \BibitemShut
  {NoStop}%
\bibitem [{\citenamefont {Davis}\ \emph {et~al.}(2012)\citenamefont {Davis},
  \citenamefont {Lim}, \citenamefont {Sakstein},\ and\ \citenamefont
  {Shaw}}]{Davis:2011qf}%
  \BibitemOpen
  \bibfield  {author} {\bibinfo {author} {\bibfnamefont {A.-C.}\ \bibnamefont
  {Davis}}, \bibinfo {author} {\bibfnamefont {E.~A.}\ \bibnamefont {Lim}},
  \bibinfo {author} {\bibfnamefont {J.}~\bibnamefont {Sakstein}}, \ and\
  \bibinfo {author} {\bibfnamefont {D.}~\bibnamefont {Shaw}},\ }\href {\doibase
  10.1103/PhysRevD.85.123006} {\bibfield  {journal} {\bibinfo  {journal}
  {Phys.Rev.}\ }\textbf {\bibinfo {volume} {D85}},\ \bibinfo {pages} {123006}
  (\bibinfo {year} {2012})},\ \Eprint {http://arxiv.org/abs/1102.5278}
  {arXiv:1102.5278 [astro-ph.CO]} \BibitemShut {NoStop}%
\bibitem [{\citenamefont {Jain}\ \emph {et~al.}(2013)\citenamefont {Jain},
  \citenamefont {Vikram},\ and\ \citenamefont {Sakstein}}]{Jain:2012tn}%
  \BibitemOpen
  \bibfield  {author} {\bibinfo {author} {\bibfnamefont {B.}~\bibnamefont
  {Jain}}, \bibinfo {author} {\bibfnamefont {V.}~\bibnamefont {Vikram}}, \ and\
  \bibinfo {author} {\bibfnamefont {J.}~\bibnamefont {Sakstein}},\ }\href
  {\doibase 10.1088/0004-637X/779/1/39} {\bibfield  {journal} {\bibinfo
  {journal} {Astrophys.J.}\ }\textbf {\bibinfo {volume} {779}},\ \bibinfo
  {pages} {39} (\bibinfo {year} {2013})},\ \Eprint
  {http://arxiv.org/abs/1204.6044} {arXiv:1204.6044 [astro-ph.CO]} \BibitemShut
  {NoStop}%
\bibitem [{\citenamefont {Brax}\ \emph {et~al.}(2014)\citenamefont {Brax},
  \citenamefont {Davis},\ and\ \citenamefont {Sakstein}}]{Brax:2013uh}%
  \BibitemOpen
  \bibfield  {author} {\bibinfo {author} {\bibfnamefont {P.}~\bibnamefont
  {Brax}}, \bibinfo {author} {\bibfnamefont {A.-C.}\ \bibnamefont {Davis}}, \
  and\ \bibinfo {author} {\bibfnamefont {J.}~\bibnamefont {Sakstein}},\ }\href
  {\doibase 10.1088/0264-9381/31/22/225001} {\bibfield  {journal} {\bibinfo
  {journal} {Class.Quant.Grav.}\ }\textbf {\bibinfo {volume} {31}},\ \bibinfo
  {pages} {225001} (\bibinfo {year} {2014})},\ \Eprint
  {http://arxiv.org/abs/1301.5587} {arXiv:1301.5587 [gr-qc]} \BibitemShut
  {NoStop}%
\bibitem [{\citenamefont {Sakstein}(2013)}]{Sakstein:2013pda}%
  \BibitemOpen
  \bibfield  {author} {\bibinfo {author} {\bibfnamefont {J.}~\bibnamefont
  {Sakstein}},\ }\href {\doibase 10.1103/PhysRevD.88.124013} {\bibfield
  {journal} {\bibinfo  {journal} {Phys.Rev.}\ }\textbf {\bibinfo {volume}
  {D88}},\ \bibinfo {pages} {124013} (\bibinfo {year} {2013})},\ \Eprint
  {http://arxiv.org/abs/1309.0495} {arXiv:1309.0495 [astro-ph.CO]} \BibitemShut
  {NoStop}%
\bibitem [{\citenamefont {Vikram}\ \emph {et~al.}(2014)\citenamefont {Vikram},
  \citenamefont {Sakstein}, \citenamefont {Davis},\ and\ \citenamefont
  {Neil}}]{Vikram:2014uza}%
  \BibitemOpen
  \bibfield  {author} {\bibinfo {author} {\bibfnamefont {V.}~\bibnamefont
  {Vikram}}, \bibinfo {author} {\bibfnamefont {J.}~\bibnamefont {Sakstein}},
  \bibinfo {author} {\bibfnamefont {C.}~\bibnamefont {Davis}}, \ and\ \bibinfo
  {author} {\bibfnamefont {A.}~\bibnamefont {Neil}},\ }\href@noop {} {\
  (\bibinfo {year} {2014})},\ \Eprint {http://arxiv.org/abs/1407.6044}
  {arXiv:1407.6044 [astro-ph.CO]} \BibitemShut {NoStop}%
\bibitem [{\citenamefont {Sakstein}\ \emph {et~al.}(2014)\citenamefont
  {Sakstein}, \citenamefont {Jain},\ and\ \citenamefont
  {Vikram}}]{Sakstein:2014nfa}%
  \BibitemOpen
  \bibfield  {author} {\bibinfo {author} {\bibfnamefont {J.}~\bibnamefont
  {Sakstein}}, \bibinfo {author} {\bibfnamefont {B.}~\bibnamefont {Jain}}, \
  and\ \bibinfo {author} {\bibfnamefont {V.}~\bibnamefont {Vikram}},\ }\href
  {\doibase 10.1142/S0218271814420024} {\  (\bibinfo {year} {2014}),\
  10.1142/S0218271814420024},\ \Eprint {http://arxiv.org/abs/1409.3708}
  {arXiv:1409.3708 [astro-ph.CO]} \BibitemShut {NoStop}%
\bibitem [{\citenamefont {Sakstein}(2015{\natexlab{a}})}]{Sakstein:2015oqa}%
  \BibitemOpen
  \bibfield  {author} {\bibinfo {author} {\bibfnamefont {J.}~\bibnamefont
  {Sakstein}},\ }\href@noop {} {\  (\bibinfo {year} {2015}{\natexlab{a}})},\
  \Eprint {http://arxiv.org/abs/1502.04503} {arXiv:1502.04503 [astro-ph.CO]}
  \BibitemShut {NoStop}%
\bibitem [{\citenamefont {Koyama}\ and\ \citenamefont
  {Sakstein}(2015)}]{Koyama:2015oma}%
  \BibitemOpen
  \bibfield  {author} {\bibinfo {author} {\bibfnamefont {K.}~\bibnamefont
  {Koyama}}\ and\ \bibinfo {author} {\bibfnamefont {J.}~\bibnamefont
  {Sakstein}},\ }\href {\doibase 10.1103/PhysRevD.91.124066} {\bibfield
  {journal} {\bibinfo  {journal} {Phys.Rev.}\ }\textbf {\bibinfo {volume}
  {D91}},\ \bibinfo {pages} {124066} (\bibinfo {year} {2015})},\ \Eprint
  {http://arxiv.org/abs/1502.06872} {arXiv:1502.06872 [astro-ph.CO]}
  \BibitemShut {NoStop}%
\bibitem [{\citenamefont {Sakstein}\ and\ \citenamefont
  {Koyama}(2015)}]{Sakstein:2015aqx}%
  \BibitemOpen
  \bibfield  {author} {\bibinfo {author} {\bibfnamefont {J.}~\bibnamefont
  {Sakstein}}\ and\ \bibinfo {author} {\bibfnamefont {K.}~\bibnamefont
  {Koyama}},\ }\href {\doibase 10.1142/S0218271815440216} {\bibfield  {journal}
  {\bibinfo  {journal} {Int. J. Mod. Phys.}\ }\textbf {\bibinfo {volume}
  {D24}},\ \bibinfo {pages} {1544021} (\bibinfo {year} {2015})}\BibitemShut
  {NoStop}%
\bibitem [{\citenamefont {Vainshtein}(1972)}]{Vainshtein:1972sx}%
  \BibitemOpen
  \bibfield  {author} {\bibinfo {author} {\bibfnamefont {A.}~\bibnamefont
  {Vainshtein}},\ }\href {\doibase 10.1016/0370-2693(72)90147-5} {\bibfield
  {journal} {\bibinfo  {journal} {Phys.Lett.}\ }\textbf {\bibinfo {volume}
  {B39}},\ \bibinfo {pages} {393} (\bibinfo {year} {1972})}\BibitemShut
  {NoStop}%
\bibitem [{\citenamefont {Babichev}\ \emph {et~al.}(2009)\citenamefont
  {Babichev}, \citenamefont {Deffayet},\ and\ \citenamefont
  {Ziour}}]{Babichev:2009us}%
  \BibitemOpen
  \bibfield  {author} {\bibinfo {author} {\bibfnamefont {E.}~\bibnamefont
  {Babichev}}, \bibinfo {author} {\bibfnamefont {C.}~\bibnamefont {Deffayet}},
  \ and\ \bibinfo {author} {\bibfnamefont {R.}~\bibnamefont {Ziour}},\ }\href
  {\doibase 10.1088/1126-6708/2009/05/098} {\bibfield  {journal} {\bibinfo
  {journal} {JHEP}\ }\textbf {\bibinfo {volume} {0905}},\ \bibinfo {pages}
  {098} (\bibinfo {year} {2009})},\ \Eprint {http://arxiv.org/abs/0901.0393}
  {arXiv:0901.0393 [hep-th]} \BibitemShut {NoStop}%
\bibitem [{\citenamefont {Kimura}\ \emph {et~al.}(2012)\citenamefont {Kimura},
  \citenamefont {Kobayashi},\ and\ \citenamefont {Yamamoto}}]{Kimura:2011dc}%
  \BibitemOpen
  \bibfield  {author} {\bibinfo {author} {\bibfnamefont {R.}~\bibnamefont
  {Kimura}}, \bibinfo {author} {\bibfnamefont {T.}~\bibnamefont {Kobayashi}}, \
  and\ \bibinfo {author} {\bibfnamefont {K.}~\bibnamefont {Yamamoto}},\ }\href
  {\doibase 10.1103/PhysRevD.85.024023} {\bibfield  {journal} {\bibinfo
  {journal} {Phys.Rev.}\ }\textbf {\bibinfo {volume} {D85}},\ \bibinfo {pages}
  {024023} (\bibinfo {year} {2012})},\ \Eprint {http://arxiv.org/abs/1111.6749}
  {arXiv:1111.6749 [astro-ph.CO]} \BibitemShut {NoStop}%
\bibitem [{\citenamefont {Koyama}\ \emph {et~al.}(2013)\citenamefont {Koyama},
  \citenamefont {Niz},\ and\ \citenamefont {Tasinato}}]{Koyama:2013paa}%
  \BibitemOpen
  \bibfield  {author} {\bibinfo {author} {\bibfnamefont {K.}~\bibnamefont
  {Koyama}}, \bibinfo {author} {\bibfnamefont {G.}~\bibnamefont {Niz}}, \ and\
  \bibinfo {author} {\bibfnamefont {G.}~\bibnamefont {Tasinato}},\ }\href
  {\doibase 10.1103/PhysRevD.88.021502} {\bibfield  {journal} {\bibinfo
  {journal} {Phys.Rev.}\ }\textbf {\bibinfo {volume} {D88}},\ \bibinfo {pages}
  {021502} (\bibinfo {year} {2013})},\ \Eprint {http://arxiv.org/abs/1305.0279}
  {arXiv:1305.0279 [hep-th]} \BibitemShut {NoStop}%
\bibitem [{\citenamefont {Kobayashi}\ \emph {et~al.}(2014)\citenamefont
  {Kobayashi}, \citenamefont {Watanabe},\ and\ \citenamefont
  {Yamauchi}}]{Kobayashi:2014ida}%
  \BibitemOpen
  \bibfield  {author} {\bibinfo {author} {\bibfnamefont {T.}~\bibnamefont
  {Kobayashi}}, \bibinfo {author} {\bibfnamefont {Y.}~\bibnamefont {Watanabe}},
  \ and\ \bibinfo {author} {\bibfnamefont {D.}~\bibnamefont {Yamauchi}},\
  }\href@noop {} {\  (\bibinfo {year} {2014})},\ \Eprint
  {http://arxiv.org/abs/1411.4130} {arXiv:1411.4130 [gr-qc]} \BibitemShut
  {NoStop}%
\bibitem [{\citenamefont {Gleyzes}\ \emph
  {et~al.}(2014{\natexlab{a}})\citenamefont {Gleyzes}, \citenamefont
  {Langlois}, \citenamefont {Piazza},\ and\ \citenamefont
  {Vernizzi}}]{Gleyzes:2014dya}%
  \BibitemOpen
  \bibfield  {author} {\bibinfo {author} {\bibfnamefont {J.}~\bibnamefont
  {Gleyzes}}, \bibinfo {author} {\bibfnamefont {D.}~\bibnamefont {Langlois}},
  \bibinfo {author} {\bibfnamefont {F.}~\bibnamefont {Piazza}}, \ and\ \bibinfo
  {author} {\bibfnamefont {F.}~\bibnamefont {Vernizzi}},\ }\href@noop {} {\
  (\bibinfo {year} {2014}{\natexlab{a}})},\ \Eprint
  {http://arxiv.org/abs/1404.6495} {arXiv:1404.6495 [hep-th]} \BibitemShut
  {NoStop}%
\bibitem [{\citenamefont {Gleyzes}\ \emph
  {et~al.}(2014{\natexlab{b}})\citenamefont {Gleyzes}, \citenamefont
  {Langlois}, \citenamefont {Piazza},\ and\ \citenamefont
  {Vernizzi}}]{Gleyzes:2014qga}%
  \BibitemOpen
  \bibfield  {author} {\bibinfo {author} {\bibfnamefont {J.}~\bibnamefont
  {Gleyzes}}, \bibinfo {author} {\bibfnamefont {D.}~\bibnamefont {Langlois}},
  \bibinfo {author} {\bibfnamefont {F.}~\bibnamefont {Piazza}}, \ and\ \bibinfo
  {author} {\bibfnamefont {F.}~\bibnamefont {Vernizzi}},\ }\href@noop {} {\
  (\bibinfo {year} {2014}{\natexlab{b}})},\ \Eprint
  {http://arxiv.org/abs/1408.1952} {arXiv:1408.1952 [astro-ph.CO]} \BibitemShut
  {NoStop}%
\bibitem [{\citenamefont {Saito}\ \emph {et~al.}(2015)\citenamefont {Saito},
  \citenamefont {Yamauchi}, \citenamefont {Mizuno}, \citenamefont {Gleyzes},\
  and\ \citenamefont {Langlois}}]{Saito:2015fza}%
  \BibitemOpen
  \bibfield  {author} {\bibinfo {author} {\bibfnamefont {R.}~\bibnamefont
  {Saito}}, \bibinfo {author} {\bibfnamefont {D.}~\bibnamefont {Yamauchi}},
  \bibinfo {author} {\bibfnamefont {S.}~\bibnamefont {Mizuno}}, \bibinfo
  {author} {\bibfnamefont {J.}~\bibnamefont {Gleyzes}}, \ and\ \bibinfo
  {author} {\bibfnamefont {D.}~\bibnamefont {Langlois}},\ }\href {\doibase
  10.1088/1475-7516/2015/06/008} {\bibfield  {journal} {\bibinfo  {journal}
  {JCAP}\ }\textbf {\bibinfo {volume} {1506}},\ \bibinfo {pages} {008}
  (\bibinfo {year} {2015})},\ \Eprint {http://arxiv.org/abs/1503.01448}
  {arXiv:1503.01448 [gr-qc]} \BibitemShut {NoStop}%
\bibitem [{\citenamefont {Sakstein}(2015{\natexlab{b}})}]{Sakstein:2015zoa}%
  \BibitemOpen
  \bibfield  {author} {\bibinfo {author} {\bibfnamefont {J.}~\bibnamefont
  {Sakstein}},\ }\href@noop {} {\  (\bibinfo {year} {2015}{\natexlab{b}})},\
  \Eprint {http://arxiv.org/abs/1510.05964} {arXiv:1510.05964 [astro-ph.CO]}
  \BibitemShut {NoStop}%
\bibitem [{\citenamefont {Esposito-Farese}(2004)}]{EspositoFarese:2004cc}%
  \BibitemOpen
  \bibfield  {author} {\bibinfo {author} {\bibfnamefont {G.}~\bibnamefont
  {Esposito-Farese}},\ }\href {\doibase 10.1063/1.1835173} {\bibfield
  {journal} {\bibinfo  {journal} {AIP Conf.Proc.}\ }\textbf {\bibinfo {volume}
  {736}},\ \bibinfo {pages} {35} (\bibinfo {year} {2004})},\ \Eprint
  {http://arxiv.org/abs/gr-qc/0409081} {arXiv:gr-qc/0409081 [gr-qc]}
  \BibitemShut {NoStop}%
\bibitem [{\citenamefont {Nicolis}\ \emph {et~al.}(2009)\citenamefont
  {Nicolis}, \citenamefont {Rattazzi},\ and\ \citenamefont
  {Trincherini}}]{Nicolis:2008in}%
  \BibitemOpen
  \bibfield  {author} {\bibinfo {author} {\bibfnamefont {A.}~\bibnamefont
  {Nicolis}}, \bibinfo {author} {\bibfnamefont {R.}~\bibnamefont {Rattazzi}}, \
  and\ \bibinfo {author} {\bibfnamefont {E.}~\bibnamefont {Trincherini}},\
  }\href {\doibase 10.1103/PhysRevD.79.064036} {\bibfield  {journal} {\bibinfo
  {journal} {Phys.Rev.}\ }\textbf {\bibinfo {volume} {D79}},\ \bibinfo {pages}
  {064036} (\bibinfo {year} {2009})},\ \Eprint {http://arxiv.org/abs/0811.2197}
  {arXiv:0811.2197 [hep-th]} \BibitemShut {NoStop}%
\bibitem [{\citenamefont {Khoury}(2013)}]{Khoury:2013tda}%
  \BibitemOpen
  \bibfield  {author} {\bibinfo {author} {\bibfnamefont {J.}~\bibnamefont
  {Khoury}},\ }\href@noop {} {\  (\bibinfo {year} {2013})},\ \Eprint
  {http://arxiv.org/abs/1312.2006} {arXiv:1312.2006 [astro-ph.CO]} \BibitemShut
  {NoStop}%
\bibitem [{\citenamefont {Bellini}\ and\ \citenamefont
  {Sawicki}(2014)}]{Bellini:2014fua}%
  \BibitemOpen
  \bibfield  {author} {\bibinfo {author} {\bibfnamefont {E.}~\bibnamefont
  {Bellini}}\ and\ \bibinfo {author} {\bibfnamefont {I.}~\bibnamefont
  {Sawicki}},\ }\href {\doibase 10.1088/1475-7516/2014/07/050} {\bibfield
  {journal} {\bibinfo  {journal} {JCAP}\ }\textbf {\bibinfo {volume} {1407}},\
  \bibinfo {pages} {050} (\bibinfo {year} {2014})},\ \Eprint
  {http://arxiv.org/abs/1404.3713} {arXiv:1404.3713 [astro-ph.CO]} \BibitemShut
  {NoStop}%
\bibitem [{\citenamefont {Zumalacárregui}\ and\ \citenamefont
  {García-Bellido}(2014)}]{Zumalacarregui:2013pma}%
  \BibitemOpen
  \bibfield  {author} {\bibinfo {author} {\bibfnamefont {M.}~\bibnamefont
  {Zumalacárregui}}\ and\ \bibinfo {author} {\bibfnamefont {J.}~\bibnamefont
  {García-Bellido}},\ }\href {\doibase 10.1103/PhysRevD.89.064046} {\bibfield
  {journal} {\bibinfo  {journal} {Phys.Rev.}\ }\textbf {\bibinfo {volume}
  {D89}},\ \bibinfo {pages} {064046} (\bibinfo {year} {2014})},\ \Eprint
  {http://arxiv.org/abs/1308.4685} {arXiv:1308.4685 [gr-qc]} \BibitemShut
  {NoStop}%
\bibitem [{\citenamefont {Gao}(2014{\natexlab{a}})}]{Gao:2014soa}%
  \BibitemOpen
  \bibfield  {author} {\bibinfo {author} {\bibfnamefont {X.}~\bibnamefont
  {Gao}},\ }\href {\doibase 10.1103/PhysRevD.90.081501} {\bibfield  {journal}
  {\bibinfo  {journal} {Phys. Rev.}\ }\textbf {\bibinfo {volume} {D90}},\
  \bibinfo {pages} {081501} (\bibinfo {year} {2014}{\natexlab{a}})},\ \Eprint
  {http://arxiv.org/abs/1406.0822} {arXiv:1406.0822 [gr-qc]} \BibitemShut
  {NoStop}%
\bibitem [{\citenamefont {Gao}(2014{\natexlab{b}})}]{Gao:2014fra}%
  \BibitemOpen
  \bibfield  {author} {\bibinfo {author} {\bibfnamefont {X.}~\bibnamefont
  {Gao}},\ }\href {\doibase 10.1103/PhysRevD.90.104033} {\bibfield  {journal}
  {\bibinfo  {journal} {Phys. Rev.}\ }\textbf {\bibinfo {volume} {D90}},\
  \bibinfo {pages} {104033} (\bibinfo {year} {2014}{\natexlab{b}})},\ \Eprint
  {http://arxiv.org/abs/1409.6708} {arXiv:1409.6708 [gr-qc]} \BibitemShut
  {NoStop}%
\bibitem [{\citenamefont {Burrows}\ and\ \citenamefont
  {Liebert}(1993)}]{Burrows:1992fg}%
  \BibitemOpen
  \bibfield  {author} {\bibinfo {author} {\bibfnamefont {A.}~\bibnamefont
  {Burrows}}\ and\ \bibinfo {author} {\bibfnamefont {J.}~\bibnamefont
  {Liebert}},\ }\href {\doibase 10.1103/RevModPhys.65.301} {\bibfield
  {journal} {\bibinfo  {journal} {Rev.Mod.Phys.}\ }\textbf {\bibinfo {volume}
  {65}},\ \bibinfo {pages} {301} (\bibinfo {year} {1993})}\BibitemShut
  {NoStop}%
\bibitem [{\citenamefont {Kippenhahn}\ and\ \citenamefont
  {Weigert}(1990)}]{kippenhahn1990stellar}%
  \BibitemOpen
  \bibfield  {author} {\bibinfo {author} {\bibfnamefont {R.}~\bibnamefont
  {Kippenhahn}}\ and\ \bibinfo {author} {\bibfnamefont {A.}~\bibnamefont
  {Weigert}},\ }\href@noop {} {\emph {\bibinfo {title} {Stellar structure and
  evolution}}},\ Astronomy and astrophysics library\ (\bibinfo  {publisher}
  {Springer},\ \bibinfo {year} {1990})\BibitemShut {NoStop}%
\bibitem [{\citenamefont {{Saumon}}\ \emph {et~al.}(1996)\citenamefont
  {{Saumon}}, \citenamefont {{Hubbard}}, \citenamefont {{Burrows}},
  \citenamefont {{Guillot}}, \citenamefont {{Lunine}},\ and\ \citenamefont
  {{Chabrier}}}]{1996ApJ...460..993S}%
  \BibitemOpen
  \bibfield  {author} {\bibinfo {author} {\bibfnamefont {D.}~\bibnamefont
  {{Saumon}}}, \bibinfo {author} {\bibfnamefont {W.~B.}\ \bibnamefont
  {{Hubbard}}}, \bibinfo {author} {\bibfnamefont {A.}~\bibnamefont
  {{Burrows}}}, \bibinfo {author} {\bibfnamefont {T.}~\bibnamefont
  {{Guillot}}}, \bibinfo {author} {\bibfnamefont {J.~I.}\ \bibnamefont
  {{Lunine}}}, \ and\ \bibinfo {author} {\bibfnamefont {G.}~\bibnamefont
  {{Chabrier}}},\ }\href {\doibase 10.1086/177027} {\bibfield  {journal}
  {\bibinfo  {journal} {\apj}\ }\textbf {\bibinfo {volume} {460}},\ \bibinfo
  {pages} {993} (\bibinfo {year} {1996})},\ \Eprint
  {http://arxiv.org/abs/astro-ph/9510046} {astro-ph/9510046} \BibitemShut
  {NoStop}%
\bibitem [{\citenamefont {Lissauer}\ and\ \citenamefont
  {de~Pater}(2013)}]{lissauer2013fundamental}%
  \BibitemOpen
  \bibfield  {author} {\bibinfo {author} {\bibfnamefont {J.}~\bibnamefont
  {Lissauer}}\ and\ \bibinfo {author} {\bibfnamefont {I.}~\bibnamefont
  {de~Pater}},\ }\href {https://books.google.co.uk/books?id=0iggAwAAQBAJ}
  {\emph {\bibinfo {title} {Fundamental Planetary Science: Physics, Chemistry
  and Habitability}}}\ (\bibinfo  {publisher} {Cambridge University Press},\
  \bibinfo {year} {2013})\BibitemShut {NoStop}%
\bibitem [{\citenamefont
  {Chandrasekhar}(2012)}]{chandrasekhar2012introduction}%
  \BibitemOpen
  \bibfield  {author} {\bibinfo {author} {\bibfnamefont {S.}~\bibnamefont
  {Chandrasekhar}},\ }\href {http://books.google.co.uk/books?id=joWn8s2BF04C}
  {\emph {\bibinfo {title} {An Introduction to the Study of Stellar
  Structure}}},\ Dover Books on Astronomy Series\ (\bibinfo  {publisher} {Dover
  Publications, Incorporated},\ \bibinfo {year} {2012})\BibitemShut {NoStop}%
\bibitem [{\citenamefont {{Kumar}}(1962)}]{1962iss..rept....1K}%
  \BibitemOpen
  \bibfield  {author} {\bibinfo {author} {\bibfnamefont {S.~S.}\ \bibnamefont
  {{Kumar}}},\ }\href@noop {} {\emph {\bibinfo {title} {Institute for Space
  Studies Report Number X-644-62-78 (1962)}}},\ \bibinfo {type} {Tech. Rep.}\
  (\bibinfo {year} {1962})\BibitemShut {NoStop}%
\bibitem [{\citenamefont {{Hayashi}}\ and\ \citenamefont
  {{Nakano}}(1963)}]{1963PThPh..30..460H}%
  \BibitemOpen
  \bibfield  {author} {\bibinfo {author} {\bibfnamefont {C.}~\bibnamefont
  {{Hayashi}}}\ and\ \bibinfo {author} {\bibfnamefont {T.}~\bibnamefont
  {{Nakano}}},\ }\href {\doibase 10.1143/PTP.30.460} {\bibfield  {journal}
  {\bibinfo  {journal} {Progress of Theoretical Physics}\ }\textbf {\bibinfo
  {volume} {30}},\ \bibinfo {pages} {460} (\bibinfo {year} {1963})}\BibitemShut
  {NoStop}%
\bibitem [{\citenamefont {Burgasser}\ and\ \citenamefont
  {Blake}(2009)}]{Burgasser:2009ym}%
  \BibitemOpen
  \bibfield  {author} {\bibinfo {author} {\bibfnamefont {A.~J.}\ \bibnamefont
  {Burgasser}}\ and\ \bibinfo {author} {\bibfnamefont {C.~H.}\ \bibnamefont
  {Blake}},\ }\href {\doibase 10.1088/0004-6256/137/6/4621} {\bibfield
  {journal} {\bibinfo  {journal} {Astron.J.}\ }\textbf {\bibinfo {volume}
  {137}},\ \bibinfo {pages} {4621} (\bibinfo {year} {2009})},\ \Eprint
  {http://arxiv.org/abs/0903.3440} {arXiv:0903.3440 [astro-ph.SR]} \BibitemShut
  {NoStop}%
\bibitem [{\citenamefont {{Graboske}}\ \emph {et~al.}(1973)\citenamefont
  {{Graboske}}, \citenamefont {{Dewitt}}, \citenamefont {{Grossman}},\ and\
  \citenamefont {{Cooper}}}]{1973ApJ...181..457G}%
  \BibitemOpen
  \bibfield  {author} {\bibinfo {author} {\bibfnamefont {H.~C.}\ \bibnamefont
  {{Graboske}}}, \bibinfo {author} {\bibfnamefont {H.~E.}\ \bibnamefont
  {{Dewitt}}}, \bibinfo {author} {\bibfnamefont {A.~S.}\ \bibnamefont
  {{Grossman}}}, \ and\ \bibinfo {author} {\bibfnamefont {M.~S.}\ \bibnamefont
  {{Cooper}}},\ }\href {\doibase 10.1086/152062} {\bibfield  {journal}
  {\bibinfo  {journal} {\apj}\ }\textbf {\bibinfo {volume} {181}},\ \bibinfo
  {pages} {457} (\bibinfo {year} {1973})}\BibitemShut {NoStop}%
\bibitem [{\citenamefont {Fowler}\ \emph {et~al.}(1975)\citenamefont {Fowler},
  \citenamefont {Caughlan},\ and\ \citenamefont {Zimmerman}}]{Fowler:1975kz}%
  \BibitemOpen
  \bibfield  {author} {\bibinfo {author} {\bibfnamefont {W.}~\bibnamefont
  {Fowler}}, \bibinfo {author} {\bibfnamefont {G.}~\bibnamefont {Caughlan}}, \
  and\ \bibinfo {author} {\bibfnamefont {B.}~\bibnamefont {Zimmerman}},\ }\href
  {\doibase 10.1146/annurev.aa.13.090175.000441} {\bibfield  {journal}
  {\bibinfo  {journal} {Ann.Rev.Astron.Astrophys.}\ }\textbf {\bibinfo {volume}
  {13}},\ \bibinfo {pages} {69} (\bibinfo {year} {1975})}\BibitemShut {NoStop}%
\bibitem [{\citenamefont {Chabrier}\ \emph {et~al.}(2009)\citenamefont
  {Chabrier}, \citenamefont {Baraffe}, \citenamefont {Leconte}, \citenamefont
  {Gallardo},\ and\ \citenamefont {barman}}]{Chabrier:2008bc}%
  \BibitemOpen
  \bibfield  {author} {\bibinfo {author} {\bibfnamefont {G.}~\bibnamefont
  {Chabrier}}, \bibinfo {author} {\bibfnamefont {I.}~\bibnamefont {Baraffe}},
  \bibinfo {author} {\bibfnamefont {J.}~\bibnamefont {Leconte}}, \bibinfo
  {author} {\bibfnamefont {J.}~\bibnamefont {Gallardo}}, \ and\ \bibinfo
  {author} {\bibfnamefont {T.}~\bibnamefont {barman}},\ }\href {\doibase
  10.1063/1.3099078} {\bibfield  {journal} {\bibinfo  {journal} {AIP
  Conf.Proc.}\ }\textbf {\bibinfo {volume} {1094}},\ \bibinfo {pages} {102}
  (\bibinfo {year} {2009})},\ \Eprint {http://arxiv.org/abs/0810.5085}
  {arXiv:0810.5085 [astro-ph]} \BibitemShut {NoStop}%
\bibitem [{\citenamefont {Hatzes}\ and\ \citenamefont
  {Rauer}(2015)}]{2041-8205-810-2-L25}%
  \BibitemOpen
  \bibfield  {author} {\bibinfo {author} {\bibfnamefont {A.~P.}\ \bibnamefont
  {Hatzes}}\ and\ \bibinfo {author} {\bibfnamefont {H.}~\bibnamefont {Rauer}},\
  }\href {http://stacks.iop.org/2041-8205/810/i=2/a=L25} {\bibfield  {journal}
  {\bibinfo  {journal} {The Astrophysical Journal Letters}\ }\textbf {\bibinfo
  {volume} {810}},\ \bibinfo {pages} {L25} (\bibinfo {year}
  {2015})}\BibitemShut {NoStop}%
\bibitem [{\citenamefont {Sozzetti}\ \emph {et~al.}(2003)\citenamefont
  {Sozzetti}, \citenamefont {Casertano}, \citenamefont {Lattanzi},\ and\
  \citenamefont {Spagna}}]{Sozzetti:2003vn}%
  \BibitemOpen
  \bibfield  {author} {\bibinfo {author} {\bibfnamefont {A.}~\bibnamefont
  {Sozzetti}}, \bibinfo {author} {\bibfnamefont {S.}~\bibnamefont {Casertano}},
  \bibinfo {author} {\bibfnamefont {M.~G.}\ \bibnamefont {Lattanzi}}, \ and\
  \bibinfo {author} {\bibfnamefont {A.}~\bibnamefont {Spagna}},\ }\bibfield
  {booktitle} {\emph {\bibinfo {booktitle} {{Toward Other Earths: Darwin / TPF
  and the Search for Extrasolar Terrestrial Planets Heidelberg, Germany, April
  22-25, 2003}}},\ }\href@noop {} {\  (\bibinfo {year} {2003})},\ \bibinfo
  {note} {[ESA Spec. Publ.539,605(2003)]},\ \Eprint
  {http://arxiv.org/abs/astro-ph/0305111} {arXiv:astro-ph/0305111 [astro-ph]}
  \BibitemShut {NoStop}%
\bibitem [{\citenamefont {{Sozzetti}}\ \emph {et~al.}(2014)\citenamefont
  {{Sozzetti}}, \citenamefont {{Giacobbe}}, \citenamefont {{Lattanzi}},
  \citenamefont {{Micela}}, \citenamefont {{Morbidelli}},\ and\ \citenamefont
  {{Tinetti}}}]{2014MNRAS.437..497S}%
  \BibitemOpen
  \bibfield  {author} {\bibinfo {author} {\bibfnamefont {A.}~\bibnamefont
  {{Sozzetti}}}, \bibinfo {author} {\bibfnamefont {P.}~\bibnamefont
  {{Giacobbe}}}, \bibinfo {author} {\bibfnamefont {M.~G.}\ \bibnamefont
  {{Lattanzi}}}, \bibinfo {author} {\bibfnamefont {G.}~\bibnamefont
  {{Micela}}}, \bibinfo {author} {\bibfnamefont {R.}~\bibnamefont
  {{Morbidelli}}}, \ and\ \bibinfo {author} {\bibfnamefont {G.}~\bibnamefont
  {{Tinetti}}},\ }\href {\doibase 10.1093/mnras/stt1899} {\bibfield  {journal}
  {\bibinfo  {journal} {MNRAS}\ }\textbf {\bibinfo {volume} {437}},\ \bibinfo
  {pages} {497} (\bibinfo {year} {2014})},\ \Eprint
  {http://arxiv.org/abs/1310.1405} {arXiv:1310.1405 [astro-ph.EP]} \BibitemShut
  {NoStop}%
\bibitem [{\citenamefont {Horedt}(2006)}]{horedt2006polytropes}%
  \BibitemOpen
  \bibfield  {author} {\bibinfo {author} {\bibfnamefont {G.}~\bibnamefont
  {Horedt}},\ }\href {https://books.google.co.uk/books?id=Hm3rBwAAQBAJ} {\emph
  {\bibinfo {title} {Polytropes: Applications in Astrophysics and Related
  Fields}}},\ Astrophysics and Space Science Library\ (\bibinfo  {publisher}
  {Springer Netherlands},\ \bibinfo {year} {2006})\BibitemShut {NoStop}%
\bibitem [{\citenamefont {{Burrows}}\ \emph {et~al.}(1989)\citenamefont
  {{Burrows}}, \citenamefont {{Hubbard}},\ and\ \citenamefont
  {{Lunine}}}]{1989ApJ...345..939B}%
  \BibitemOpen
  \bibfield  {author} {\bibinfo {author} {\bibfnamefont {A.}~\bibnamefont
  {{Burrows}}}, \bibinfo {author} {\bibfnamefont {W.~B.}\ \bibnamefont
  {{Hubbard}}}, \ and\ \bibinfo {author} {\bibfnamefont {J.~I.}\ \bibnamefont
  {{Lunine}}},\ }\href {\doibase 10.1086/167964} {\bibfield  {journal}
  {\bibinfo  {journal} {\apj}\ }\textbf {\bibinfo {volume} {345}},\ \bibinfo
  {pages} {939} (\bibinfo {year} {1989})}\BibitemShut {NoStop}%
\bibitem [{\citenamefont {Stevenson}(1991)}]{Stevenson:1991eq}%
  \BibitemOpen
  \bibfield  {author} {\bibinfo {author} {\bibfnamefont {D.}~\bibnamefont
  {Stevenson}},\ }\href {\doibase 10.1146/annurev.aa.29.090191.001115}
  {\bibfield  {journal} {\bibinfo  {journal} {Ann.Rev.Astron.Astrophys.}\
  }\textbf {\bibinfo {volume} {29}},\ \bibinfo {pages} {163} (\bibinfo {year}
  {1991})}\BibitemShut {NoStop}%
\bibitem [{\citenamefont {{Kumar}}(1963)}]{1963ApJ...137.1121K}%
  \BibitemOpen
  \bibfield  {author} {\bibinfo {author} {\bibfnamefont {S.~S.}\ \bibnamefont
  {{Kumar}}},\ }\href {\doibase 10.1086/147589} {\bibfield  {journal} {\bibinfo
   {journal} {\apj}\ }\textbf {\bibinfo {volume} {137}},\ \bibinfo {pages}
  {1121} (\bibinfo {year} {1963})}\BibitemShut {NoStop}%
\bibitem [{\citenamefont {Burrows}\ \emph {et~al.}(1997)\citenamefont
  {Burrows}, \citenamefont {Marley}, \citenamefont {Hubbard}, \citenamefont
  {Lunine}, \citenamefont {Guillot} \emph {et~al.}}]{Burrows:1997ka}%
  \BibitemOpen
  \bibfield  {author} {\bibinfo {author} {\bibfnamefont {A.}~\bibnamefont
  {Burrows}}, \bibinfo {author} {\bibfnamefont {M.}~\bibnamefont {Marley}},
  \bibinfo {author} {\bibfnamefont {W.}~\bibnamefont {Hubbard}}, \bibinfo
  {author} {\bibfnamefont {J.}~\bibnamefont {Lunine}}, \bibinfo {author}
  {\bibfnamefont {T.}~\bibnamefont {Guillot}},  \emph {et~al.},\ }\href
  {\doibase 10.1086/305002} {\bibfield  {journal} {\bibinfo  {journal}
  {Astrophys.J.}\ }\textbf {\bibinfo {volume} {491}},\ \bibinfo {pages} {856}
  (\bibinfo {year} {1997})},\ \Eprint {http://arxiv.org/abs/astro-ph/9705201}
  {arXiv:astro-ph/9705201 [astro-ph]} \BibitemShut {NoStop}%
\bibitem [{\citenamefont {Coppenbarger}\ \emph {et~al.}(1994)\citenamefont
  {Coppenbarger}, \citenamefont {Henry},\ and\ \citenamefont
  {McCarthy}}]{Coppenbarger:1994us}%
  \BibitemOpen
  \bibfield  {author} {\bibinfo {author} {\bibfnamefont {D.~S.}\ \bibnamefont
  {Coppenbarger}}, \bibinfo {author} {\bibfnamefont {T.~J.}\ \bibnamefont
  {Henry}}, \ and\ \bibinfo {author} {\bibfnamefont {J.}~\bibnamefont
  {McCarthy}, \bibfnamefont {Donald~W.}},\ }\href {\doibase 10.1086/116966}
  {\bibfield  {journal} {\bibinfo  {journal} {Astron.J.}\ }\textbf {\bibinfo
  {volume} {107}},\ \bibinfo {pages} {1551} (\bibinfo {year}
  {1994})}\BibitemShut {NoStop}%
\bibitem [{\citenamefont {{Barbieri}}\ \emph {et~al.}(1996)\citenamefont
  {{Barbieri}}, \citenamefont {{De Marchi}}, \citenamefont {{Nota}},
  \citenamefont {{Corrain}}, \citenamefont {{Hack}}, \citenamefont
  {{Ragazzoni}},\ and\ \citenamefont {{Macchetto}}}]{1996A&A...315..418B}%
  \BibitemOpen
  \bibfield  {author} {\bibinfo {author} {\bibfnamefont {C.}~\bibnamefont
  {{Barbieri}}}, \bibinfo {author} {\bibfnamefont {G.}~\bibnamefont {{De
  Marchi}}}, \bibinfo {author} {\bibfnamefont {A.}~\bibnamefont {{Nota}}},
  \bibinfo {author} {\bibfnamefont {G.}~\bibnamefont {{Corrain}}}, \bibinfo
  {author} {\bibfnamefont {W.}~\bibnamefont {{Hack}}}, \bibinfo {author}
  {\bibfnamefont {R.}~\bibnamefont {{Ragazzoni}}}, \ and\ \bibinfo {author}
  {\bibfnamefont {D.}~\bibnamefont {{Macchetto}}},\ }\href@noop {} {\bibfield
  {journal} {\bibinfo  {journal} {Astron.Astrophys.}\ }\textbf {\bibinfo
  {volume} {315}},\ \bibinfo {pages} {418} (\bibinfo {year}
  {1996})}\BibitemShut {NoStop}%
\bibitem [{\citenamefont {Delfosse}\ \emph {et~al.}(2000)\citenamefont
  {Delfosse}, \citenamefont {Forveille}, \citenamefont {Segransan},
  \citenamefont {Beuzit}, \citenamefont {Udry} \emph
  {et~al.}}]{Delfosse:2000jr}%
  \BibitemOpen
  \bibfield  {author} {\bibinfo {author} {\bibfnamefont {X.}~\bibnamefont
  {Delfosse}}, \bibinfo {author} {\bibfnamefont {T.}~\bibnamefont {Forveille}},
  \bibinfo {author} {\bibfnamefont {D.}~\bibnamefont {Segransan}}, \bibinfo
  {author} {\bibfnamefont {J.-L.}\ \bibnamefont {Beuzit}}, \bibinfo {author}
  {\bibfnamefont {S.}~\bibnamefont {Udry}},  \emph {et~al.},\ }\href@noop {}
  {\bibfield  {journal} {\bibinfo  {journal} {Astron.Astrophys.}\ }\textbf
  {\bibinfo {volume} {364}},\ \bibinfo {pages} {217} (\bibinfo {year}
  {2000})},\ \Eprint {http://arxiv.org/abs/astro-ph/0010586}
  {arXiv:astro-ph/0010586 [astro-ph]} \BibitemShut {NoStop}%
\bibitem [{\citenamefont {Martinache}\ \emph {et~al.}(2007)\citenamefont
  {Martinache}, \citenamefont {Lloyd}, \citenamefont {Ireland}, \citenamefont
  {Yamada},\ and\ \citenamefont {Tuthill}}]{Martinache:2006ud}%
  \BibitemOpen
  \bibfield  {author} {\bibinfo {author} {\bibfnamefont {F.}~\bibnamefont
  {Martinache}}, \bibinfo {author} {\bibfnamefont {J.~P.}\ \bibnamefont
  {Lloyd}}, \bibinfo {author} {\bibfnamefont {M.~J.}\ \bibnamefont {Ireland}},
  \bibinfo {author} {\bibfnamefont {R.~S.}\ \bibnamefont {Yamada}}, \ and\
  \bibinfo {author} {\bibfnamefont {P.~G.}\ \bibnamefont {Tuthill}},\ }\href
  {\doibase 10.1086/513868} {\bibfield  {journal} {\bibinfo  {journal}
  {Astrophys.J.}\ }\textbf {\bibinfo {volume} {661}},\ \bibinfo {pages} {496}
  (\bibinfo {year} {2007})},\ \Eprint {http://arxiv.org/abs/astro-ph/0612138}
  {arXiv:astro-ph/0612138 [astro-ph]} \BibitemShut {NoStop}%
\bibitem [{\citenamefont {Hamilton}(2012)}]{hamilton2012our}%
  \BibitemOpen
  \bibfield  {author} {\bibinfo {author} {\bibfnamefont {T.}~\bibnamefont
  {Hamilton}},\ }\href {https://books.google.co.uk/books?id=TjaAxSKPxR8C}
  {\emph {\bibinfo {title} {Our Neighbor Stars: Including Brown Dwarfs}}},\
  EBL-Schweitzer\ (\bibinfo  {publisher} {Publish on Demand Global LLC},\
  \bibinfo {year} {2012})\BibitemShut {NoStop}%
\bibitem [{\citenamefont {Segransan}\ \emph {et~al.}(2000)\citenamefont
  {Segransan}, \citenamefont {Delfosse}, \citenamefont {Forveille},
  \citenamefont {Beuzit}, \citenamefont {Udry}, \citenamefont {Perrier},\ and\
  \citenamefont {Mayor}}]{Segransan:2000jq}%
  \BibitemOpen
  \bibfield  {author} {\bibinfo {author} {\bibfnamefont {D.}~\bibnamefont
  {Segransan}}, \bibinfo {author} {\bibfnamefont {X.}~\bibnamefont {Delfosse}},
  \bibinfo {author} {\bibfnamefont {T.}~\bibnamefont {Forveille}}, \bibinfo
  {author} {\bibfnamefont {J.~L.}\ \bibnamefont {Beuzit}}, \bibinfo {author}
  {\bibfnamefont {S.}~\bibnamefont {Udry}}, \bibinfo {author} {\bibfnamefont
  {C.}~\bibnamefont {Perrier}}, \ and\ \bibinfo {author} {\bibfnamefont
  {M.}~\bibnamefont {Mayor}},\ }\href@noop {} {\bibfield  {journal} {\bibinfo
  {journal} {Astron. Astrophys.}\ }\textbf {\bibinfo {volume} {364}},\ \bibinfo
  {pages} {665} (\bibinfo {year} {2000})},\ \Eprint
  {http://arxiv.org/abs/astro-ph/0010585} {arXiv:astro-ph/0010585 [astro-ph]}
  \BibitemShut {NoStop}%
\bibitem [{\citenamefont {{Salpeter}}(1992)}]{1992ApJ...393..258S}%
  \BibitemOpen
  \bibfield  {author} {\bibinfo {author} {\bibfnamefont {E.~E.}\ \bibnamefont
  {{Salpeter}}},\ }\href {\doibase 10.1086/171502} {\bibfield  {journal}
  {\bibinfo  {journal} {\apj}\ }\textbf {\bibinfo {volume} {393}},\ \bibinfo
  {pages} {258} (\bibinfo {year} {1992})}\BibitemShut {NoStop}%
\bibitem [{\citenamefont {{Kippenhahn}}(1970)}]{1970A&A.....8...50K}%
  \BibitemOpen
  \bibfield  {author} {\bibinfo {author} {\bibfnamefont {R.}~\bibnamefont
  {{Kippenhahn}}},\ }\href@noop {} {\bibfield  {journal} {\bibinfo  {journal}
  {Astron.Astrophys.}\ }\textbf {\bibinfo {volume} {8}},\ \bibinfo {pages} {50}
  (\bibinfo {year} {1970})}\BibitemShut {NoStop}%
\bibitem [{\citenamefont {{Henry}}\ and\ \citenamefont
  {{McCarthy}}(1993)}]{1993AJ....106..773H}%
  \BibitemOpen
  \bibfield  {author} {\bibinfo {author} {\bibfnamefont {T.~J.}\ \bibnamefont
  {{Henry}}}\ and\ \bibinfo {author} {\bibfnamefont {D.~W.}\ \bibnamefont
  {{McCarthy}}, \bibfnamefont {Jr.}},\ }\href {\doibase 10.1086/116685}
  {\bibfield  {journal} {\bibinfo  {journal} {Astron.J.}\ }\textbf {\bibinfo
  {volume} {106}},\ \bibinfo {pages} {773} (\bibinfo {year}
  {1993})}\BibitemShut {NoStop}%
\bibitem [{\citenamefont {Sakstein}(2014)}]{Sakstein:2014isa}%
  \BibitemOpen
  \bibfield  {author} {\bibinfo {author} {\bibfnamefont {J.}~\bibnamefont
  {Sakstein}},\ }\href {\doibase 10.1088/1475-7516/2014/12/012} {\bibfield
  {journal} {\bibinfo  {journal} {JCAP}\ }\textbf {\bibinfo {volume} {1412}},\
  \bibinfo {pages} {012} (\bibinfo {year} {2014})},\ \Eprint
  {http://arxiv.org/abs/1409.1734} {arXiv:1409.1734 [astro-ph.CO]} \BibitemShut
  {NoStop}%
\bibitem [{\citenamefont {Ip}\ \emph {et~al.}(2015)\citenamefont {Ip},
  \citenamefont {Sakstein},\ and\ \citenamefont {Schmidt}}]{Ip:2015qsa}%
  \BibitemOpen
  \bibfield  {author} {\bibinfo {author} {\bibfnamefont {H.~Y.}\ \bibnamefont
  {Ip}}, \bibinfo {author} {\bibfnamefont {J.}~\bibnamefont {Sakstein}}, \ and\
  \bibinfo {author} {\bibfnamefont {F.}~\bibnamefont {Schmidt}},\ }\href
  {\doibase 10.1088/1475-7516/2015/10/051} {\bibfield  {journal} {\bibinfo
  {journal} {JCAP}\ }\textbf {\bibinfo {volume} {1510}},\ \bibinfo {pages}
  {051} (\bibinfo {year} {2015})},\ \Eprint {http://arxiv.org/abs/1507.00568}
  {arXiv:1507.00568 [gr-qc]} \BibitemShut {NoStop}%
\bibitem [{\citenamefont {Sakstein}\ and\ \citenamefont
  {Verner}(2015)}]{Sakstein:2015jca}%
  \BibitemOpen
  \bibfield  {author} {\bibinfo {author} {\bibfnamefont {J.}~\bibnamefont
  {Sakstein}}\ and\ \bibinfo {author} {\bibfnamefont {S.}~\bibnamefont
  {Verner}},\ }\href@noop {} {\  (\bibinfo {year} {2015})},\ \Eprint
  {http://arxiv.org/abs/1509.05679} {arXiv:1509.05679 [gr-qc]} \BibitemShut
  {NoStop}%
\bibitem [{\citenamefont {Khoury}\ and\ \citenamefont
  {Weltman}(2004)}]{Khoury:2003rn}%
  \BibitemOpen
  \bibfield  {author} {\bibinfo {author} {\bibfnamefont {J.}~\bibnamefont
  {Khoury}}\ and\ \bibinfo {author} {\bibfnamefont {A.}~\bibnamefont
  {Weltman}},\ }\href {\doibase 10.1103/PhysRevD.69.044026} {\bibfield
  {journal} {\bibinfo  {journal} {Phys. Rev.}\ }\textbf {\bibinfo {volume}
  {D69}},\ \bibinfo {pages} {044026} (\bibinfo {year} {2004})},\ \Eprint
  {http://arxiv.org/abs/astro-ph/0309411} {arXiv:astro-ph/0309411} \BibitemShut
  {NoStop}%
\bibitem [{\citenamefont {Will}(1993)}]{will1993theory}%
  \BibitemOpen
  \bibfield  {author} {\bibinfo {author} {\bibfnamefont {C.}~\bibnamefont
  {Will}},\ }\href {http://books.google.co.uk/books?id=BhnUITA7sDIC} {\emph
  {\bibinfo {title} {Theory and Experiment in Gravitational Physics}}}\
  (\bibinfo  {publisher} {Cambridge University Press},\ \bibinfo {year}
  {1993})\BibitemShut {NoStop}%
\bibitem [{\citenamefont {Hui}\ \emph {et~al.}(2009)\citenamefont {Hui},
  \citenamefont {Nicolis},\ and\ \citenamefont {Stubbs}}]{Hui:2009kc}%
  \BibitemOpen
  \bibfield  {author} {\bibinfo {author} {\bibfnamefont {L.}~\bibnamefont
  {Hui}}, \bibinfo {author} {\bibfnamefont {A.}~\bibnamefont {Nicolis}}, \ and\
  \bibinfo {author} {\bibfnamefont {C.}~\bibnamefont {Stubbs}},\ }\href
  {\doibase 10.1103/PhysRevD.80.104002} {\bibfield  {journal} {\bibinfo
  {journal} {Phys.Rev.}\ }\textbf {\bibinfo {volume} {D80}},\ \bibinfo {pages}
  {104002} (\bibinfo {year} {2009})},\ \Eprint {http://arxiv.org/abs/0905.2966}
  {arXiv:0905.2966 [astro-ph.CO]} \BibitemShut {NoStop}%
\end{thebibliography}%

%

\end{document}